\documentclass[draft]{agujournal2019}
\usepackage{url} 
\usepackage{soul}
\usepackage{amsmath}
\usepackage{xcolor}
\newcommand{\rocke}{ROCKE-3D}

\draftfalse

\journalname{JGR: Planets}

\begin{document}

\title{Venusian Habitable Climate Scenarios: Modeling Venus through time and applications to slowly rotating Venus-Like Exoplanets}

\authors{M.J. Way\affil{1,2,3}, and Anthony D. Del Genio\affil{1}}

\affiliation{1}{NASA Goddard Institute for Space Studies, 2880 Broadway, New York, NY, 10025, USA}
\affiliation{2}{GSFC Sellers Exoplanet Environments Collaboration}
\affiliation{3}{Theoretical Astrophysics, Department of Physics and Astronomy, Uppsala University, Uppsala, SE-75120, Sweden}

\correspondingauthor{M.J. Way}{Michael.J.Way@nasa.gov}


\begin{keypoints}
\item Venus could have had habitable conditions for nearly 3 billion years.
\item Surface liquid water is required for any habitable scenario.
\item Solar insolation through time is not a crucial factor if a carbonate-silicate cycle is in action.
\end{keypoints}

\begin{abstract}
One popular view of Venus' climate history describes a world that has spent
much of its life with surface liquid water, plate tectonics, and a stable
temperate climate. Part of the basis for this optimistic scenario is the high
deuterium to hydrogen ratio from the Pioneer Venus mission that was interpreted
to imply Venus had a shallow ocean's worth of water throughout much of its history.
Another view is that Venus had a long lived ($\sim$ 100 million year) primordial
magma ocean with a CO$_2$ and steam atmosphere. Venus' long lived steam atmosphere
would sufficient time to dissociate most of the water vapor, allow
significant hydrogen escape and 
oxidize the magma ocean. A third scenario is that Venus had surface water and
habitable conditions early in its history for a short period of time ($<1$Gyr), but
that a moist/runaway greenhouse took effect because of a gradually warming sun, leaving
the planet desiccated ever since. Using a general circulation model we
demonstrate the viability of the first scenario using the few
observational constraints available. We further speculate that Large Igneous
Provinces and the global resurfacing 100s of millions of years ago played key roles
in ending the clement period in its history and presenting the Venus we see today.
The results have implications for what astronomers term ``the habitable zone," 
and if Venus-like exoplanets exist with clement conditions akin to modern Earth we
propose to place them in what we term the ``optimistic Venus zone."
\end{abstract}


\section*{Plain Language Summary}

We have little data on our neighbor Venus to help us understand its
climate history. Yet Earth and Venus are sister worlds: they
initially formed close to one another, and have nearly the same mass and
radius. Despite the differences in their current atmospheres and
surface temperatures, they likely have similar bulk compositions, making
comparison between them extremely valuable for illuminating their distinct
climate histories. We analyze our present data on Venus
alongside knowledge about Earth's climate history to make a number
of exciting claims. Evaluating several snapshots in time over the past 4+
billion years, we show that Venus could have sustained liquid water and
moderate temperatures for most of this period. Cloud feedbacks from a slowly
rotating world with surface liquid water reservoirs were the keys to keeping
the planet clement. Contrast this with its current surface temperature of 450
degrees and an atmosphere dominated by Carbon Dioxide and Nitrogen. Our results
demonstrate that it was not the gradual warming of the sun over the eons that
contributed to Venus’ present hot-house state. Rather we speculate that large
igneous provinces and the global resurfacing 100s of millions of years ago played  key roles in ending the clement period in its history.

\section{Introduction}\label{sec:intro}
The case for prolonged habitability of Venus has been made by a number of authors in recent years \cite<e.g.,>[]{GrinspoonBullock2007,Way2016}. If so, then if habitability was widespread and persisted over geological timescales (e.g. 10$^{9}$ years), it is possible that organisms were capable of filling a large variety of environmental niches as occurred on Earth via evolutionary processes. This has led to speculation about possible remaining life-friendly niches \cite<e.g.,>[]{MorowitzSagan1967,Grinspoon1997,Cockell1999,Schulze-Makuch2002,Schulze-Makuch2004,Dartnell2015,Limaye2017}. These possibilities alone provide sufficient motivation to understand whether early Venus was habitable or not. However, Venus is also interesting from the perspective of the large number of rocky exoplanets discovered to date inside the traditional inner edge of the habitable zone but far enough from their host stars to maintain an atmosphere \cite{Kane2019}. If some of these planets in close proximity to their host stars have long periods of habitability it may overturn traditional notions of the habitable zone  \cite<e.g.,>[]{Kopparapu2013} and influence target selection for characterization of the atmospheres of these planets.
Hence, what appears to be a modern Venus-like world close to its parent star might host surface liquid water. We refer to such habitable worlds as residing in an ``Optimistic Venus Zone," a subset of the planets in the ``Venus Zone" described in \citeA{Kane2014}. 

There are three primary requirements for the early habitability of Venus. The first is that temperatures were low enough that liquid water was capable of condensing on the surface of Venus. The second is that Venus had a sufficient inventory of water to create the conditions believed necessary for the rise of life on terrestrial worlds \cite<e.g.,>[]{Brack2010}. Third, volatile cycling and a geologically active surface must exist to regulate the climate as possibly supported by the work of \citeA{Noack2012}. For surficial water reservoirs most research has focused on the poorly constrained measurement of the D/H ratio of $\sim$ 150$\pm$30 (times Earth's value) by the Pioneer Venus Mission\newline (https://nssdc.gsfc.nasa.gov/planetary/pioneer\_venus.html) \cite{Donahue1982,Donahue1997}. A number of other ground based measurements have been made \cite{DeBergh1991,Marcq2006,Bjoraker1992,Fedorova2008} that have generally revealed values of D/H greater than 100. For a recent review of D/H measurements of Venus see Section 6.3 of \citeA{Marcq2017}.
We review the work of authors who have considered Venus' climate evolution below.

The first simple gray radiative transfer calculations of Venus' climate history suggested an early runaway greenhouse effect \cite<e.g.,>[]{Ingersoll1969,Rasool1970}. Any water would have remained in vapor form throughout Venus' early history until its loss to space via photodissociation processes \cite{GoodyWalker1972,Walker1975}. 
CO$_2$ would continue to be outgassed and would accumulate (offset by atmospheric loss over the last 4.5Gyr) to the values we see today.

\citeA{Pollack1971} used a similar model but with non-gray radiative transfer. This was probably the first work to show that Venus could have hosted liquid water on its surface in its early history, but they also demonstrated that a runaway greenhouse was possible as well. The difference between the two outcomes was \emph{mostly} due to to the specified cloud cover and water vapor content of the atmosphere. With 50\% cloud fraction the result was usually a runaway greenhouse. For an N$_2$ dominated atmosphere with a modern Earth water vapor profile, 100\% cloud cover, and a 30\% less luminous sun (akin to $\sim$4.5Ga) he achieved surface temperatures $\sim$300K. These purely radiative models did not include convection, which would limit the water vapor mixing ratio by precipitation and limit the lapse rate of temperature to the moist adiabatic value \cite{RampinoCaldeira1994} and lessen the strength of the greenhouse effect to some degree.  Likewise, they did not include an atmospheric general circulation that would determine the cloud fraction self-consistently. \citeA{DonahuePollack1983} replicated some aspects of the \citeA{Pollack1971} work and again showed that clement conditions were possible in Venus' early history.

\citeA{Kasting1984} used a 1-D radiative-convective model to demonstrate that a planet with an insolation S0X=1.45 times that of modern Earth (1973 W m$^2$ $\sim$ 3.8Gya at Venus' orbit) would have temperatures $\sim$100$^\circ$ C for a 2 bar N$_2$ H$_2$O atmosphere. This atmosphere contained a wet stratosphere, and thus photodissociation of H$_2$O and subsequent escape of hydrogen \cite<e.g.,>[]{GoodyWalker1972} could explain the long-term loss of Venus' primordial ocean and its present dry conditions. In a subsequent study \citeA{Kasting1988} claimed that clouds would cool the atmosphere enough to keep it in a moist greenhouse state, rather than the runaway state. At the same time the \citeA{Kasting1988} surface temperature for his ``Early Venus" (Figure 7) cloud-free model was $\sim$500K and hovered just at the margin of the moist vs. runaway greenhouse states. Thus it was clear early on that maintaining liquid water on ancient Venus required high fractional cloud cover, but whether physical mechanisms exist to produce these cloudy conditions was not addressed. These early Venus habitability scenarios would
have taken place within the first billion years of the planet's evolution limiting
the possibility for complex life to evolve. Venus would have subsequently entered a moist/runaway greenhouse and left us with the state it is in today. For the curious reader \citeA{BullockGrinspoon1999} provide a nice review of the literature on the possibility of an early habitable Venus. Note that in a previous work \cite{Way2016} it was shown that if habitable conditions were possible in early Venus' history that it likely remained so and that increasing solar insolation through time is not a deciding factor. In the present work we find the same, and hence believe that limited (in time) early habitability models are not supported by our simulations.

Grinspoon and Bullock, in a number of abstracts \cite<e.g.,>[]{GrinspoonBullock2003,Grinspoon2008}, described a Venus climate evolution scenario with long-lived surface habitability consistent with that described later in this paper. This work was never published, though, so the details of their calculations and the justifications for their conjectures are not documented.

The first three-dimensional (3-D) General Circulation Model (GCM) exploration of issues relevant to ancient Venus was that of \citeA{Abe2011}. Their study focused on land planets (analogous to the planet Arrakis of \emph{Dune}) with no permanent surface water reservoirs, only limited fixed amounts of ground water. In some scenarios they found that ``in principle Venus could have been a habitable land planet as recently as 1 billion years ago." Their conclusions resulted from the limited water vapor in the atmosphere caused by the modest subsurface reservoir, its limited greenhouse effect, and the albedo of the planet. Thereafter \citeA{Leconte2013} used a 3-D GCM to look at climate scenarios for Gl581 c and HD 85512 b. HD 85512 b orbits a K-dwarf with a {\emph synchronous} rotation period of $\sim$ 58 days and receives approximately 1.86 times the insolation of present day Earth. This is slightly less than present day Venus' insolation (1.9), but its host star is a K-dwarf rather than a G-dwarf and hence the peak of its Planck blackbody spectrum is at longer wavelengths. They state, ``if not for its thick atmosphere, Venus' climate would be very close to the one of HD 85512 b." They modeled a dry planet akin to that of \citeA{Abe2011} using an N$_2$ dominated atmosphere with 376ppm CO$_2$, and no water vapor. The model produced large temperate regions for a number of different atmospheric pressures (See \citeA{Leconte2013} Figures 1, 2) for their land planet setup, but they also found similar behavior in their `collapse' scenario with different atmospheric water vapor profile amounts.

A rationale for high albedo cloud cover on ancient Venus was first presented by \citeA{Yang2014}. \citeA{Yang2014} was a large parameter study looking at the inner edge of the habitable zone around solar type stars using The National Center for Atmospheric Research (NCAR) Community Atmosphere Model (CAM) with comprehensive atmospheric physics but a thermodynamic ocean. A thermodynamic ocean, also known as a slab or mixed-layer ocean, is typically of limited depth ($<$100 meters) with a prescribed horizontal ocean heat transport, or no horizontal heat transport at all. The latter implies that the temperature of each ocean grid cell is determined solely by the atmosphere and incident sunlight directly above it.  The shallow depth reduces the time lag between solar forcing and ocean response.  See \citeA{Way2017} Section 2.2.2 for details.  In essence they stepped the sidereal rotation rate of an Earth-like world from 1 to 256 sidereal days. At the same time they increased the insolation from that of modern Earth to as much as 2.6 times the modern Earth for their most slowly rotating world of 256 sidereal days period. One of their key conclusions was that the slowest rotators would have had a day-night general circulation that would generate an optically thick contiguous cloud bank in the substellar region. This cloud deck would greatly increase the planetary albedo, keeping the surface temperature moderate even for quite high values of insolation. These results were later confirmed with a completely different 3-D GCM with a fully-coupled dynamic ocean \cite{Way2018}. \citeA{Yang2014} also included a simulation with modern Venus orbital parameters, spin rate, insolation while using modern Earth topography and land/ocean mask, but again with a thermodynamic ocean. These studies provide a possible rationale for the cloud cover needed to produce temperate surface conditions
as first postulated by \citeA{Pollack1971}. Finally \citeA<>[hereafter Paper I]{Way2016} took things one step further by exploring different topographies, insolations and rotation rates to put tighter constraints on possible habitable conditions for ancient Venus.

In this paper, we extend those parameter studies to consider a wider variety of planets and attempt to justify our modeling assumptions in light of possible scenarios for Venus' evolution.  Sections \ref{section:atmosphere}--\ref{sec:evolution} review existing observational constraints and hypotheses about the composition, thickness, and evolution of Venus' atmosphere and water history; its surface and interior; and its rotation and obliquity. In Section \ref{sec:methods} we draw upon this information to inform a series of 3-D global climate model simulations to illustrate possible scenarios for an early habitable Venus that transitioned to its current inhospitable state relatively late in its history.  We discuss the implications of our results for the design of future missions to Venus and for the potential habitability of exoplanets inside the inner edge of the traditional ``habitable zone" in Section \ref{sec:results}. Finally, recently published complimentary work by \citeA{WellerKiefer2019} 
supports many of our conclusions.

\section{Atmospheric Composition and Pressure}\label{section:atmosphere}

Assuming that Venus and Earth formed from the same parts of the protoplanetary disk and thus with similar compositions \cite<e.g.,>[]{Raymond2004}, we are guided by the history of Earth whose early atmosphere was likely CO$_2$-rich and possibly cool, but not frozen due to the faint young sun for the late Hadean and early Archean (4.2-4.0 Ga) \cite<e.g.,>[]{Owen1979,Kasting1993a,ZahnleSleep2002,Valley2002,Zahnle2006,ZahnleSchaeferFegley2010,Kunze2014,CatlingKasting2017,Krissansen-Totton2018,Mello2019,Mojzsis2019}.
The picture of a ``Cool Early Earth" promoted in the early 2000s by \citeA{Valley2002,Valley2005} and with more recent zircon data \cite{Valley2014} may also be applicable to Venus' early evolutionary history if Venus survived its magma ocean (MO) phase with some liquid surface water. \citeA{Chassefiere2012} was probably the first to make the comparison of Hadean Earth with Venus, calling his  hypothesis a ``cool early Venus." 


The prospects for a cool early Earth and Venus hypothesis have improved in recent years, while the likelihood for their end in a cataclysmic ``late instability" known as the Late Heavy Bombardment (LHB) may have started to fall out of favor. 
The LHB is important because its intensity was imagined so intense that the surface temperatures of Earth would return to those before the cool early Earth period, perhaps evaporating the oceans and/or raising surface temperatures to values exceeding 100$^\circ$C. However, in recent years a number of studies have started to question the timing and strength of the LHB \cite<e.g.>[]{Boehnke2016,Zellner2017,Morbidelli2018,Quarles2019,Mojzsis2019}. For a brief overview of the latest work on the LHB see \citeA{Mann2018} and \citeA{Voosen2020}. An ``early instability" has started to gain favor in what is termed an ``accretion tail" scenario. This accretion tail scenario has recently been promoted in the works of \citeA<>[Figure 1]{Morbidelli2018} and \citeA<>[Figure 1]{Mojzsis2019}. 
In these works a late or early heavy bombardment is triggered by a period of orbital `instability' experienced by the giant planets as described in \citeA{Ribeiro2020}.
More recent work by \citeA{Ribeiro2020} demonstrates that the instability has a median timescale of 36.78-–61.5$Myr$ and is within 136$Myr$ in 75\% of their cases.
\citeA{Ribeiro2020} say the timing of an early instability fits in nicely with the survival of the Patroclus--Menoetius Jupiter Trojan as a primordial binary from the Kuiper Belt \cite{Nesvorny2018b} and is a good match to other solar system properties \cite<e.g.>[]{Clement2019}, but they explicitly discuss the many limitations of their model. In the spirit of an early instability and accretion tail scenario, Figure 4 of \citeA{Mojzsis2019} suggests that Venus accreted less than 0.01 wt\% between 4.3 and 4.1Ga, about the same as Earth. As \citeA{Mojzsis2019} notes, ``Results show that an abating impact flux from late accretion is inadequate to sterilize the surface zone."  These works demonstrate the possibility that the cool early Earth continued through the originally proposed time period of the LHB from $\sim$ 3.8--4.1Ga.
Hence we will begin our most ancient Venus simulations with CO$_2$ dominated atmospheres at $\sim$4.2Ga (during the \citeA{Valley2002} cool early Earth period) that evolves over time to N$_2$ dominated atmospheres. If the LHB were a real event we would want to begin our simulations at 3.8Ga, rather than 4.2Ga, as others have speculated previously \cite<e.g.,>[]{RampinoCaldeira1994,Lammer2018}. There are still many unanswered questions regarding the early post-MO history of the atmospheres of Venus, Earth and Mars. Even though we have far more information to discern Earth's early post-MO atmosphere it remains a complicated story yet to be fully resolved \cite<e.g.,>[]{Hirschmann2012,Nikolaou2019} and may depend on atmospheric pressure more than previously assumed \cite<e.g.,>[]{Gaillard2014,Bower2019}. 

The carbonate-silicate cycle \cite<e.g.,>[]{Walker1981,Stewart2019} is the key to keeping most of Earth's CO$_2$ locked up in rocks for much of its history rather than in the atmosphere as on present day Venus. The carbonate-silicate cycle on Earth functions via subductive-type plate tectonics, the presence of a hydrosphere and continental crust. For Earth there are several lines of evidence to suggest these may go back to the Hadean \cite<e.g.,>[and references therein]{Mojzsis2001,Hopkins2008,Harrison2009,Korenaga2013,Harrison2017,ONeill2017,Rozel2017,Kusky2018,Korenaga2018,Maruyama2018}. \citeA{Dehant2019} reviews the literature for a later beginning of plate tectonics on Earth (Section 3.2). The requirements for how plate tectonics begins are still not fully understood and hence remains an active area of research \cite<e.g.,>[]{Lenardic2019}.
If Venus had a similar early atmospheric and interior evolution to that of Earth then early volatile cycling via some form of plate tectonics is a viable hypothesis. In fact recent work demonstrates convection regimes like that of plate tectonics in Venus' recent history \cite{Gillmann2014,Davaille2017}. \citeA{Dehant2019} reviews plate tectonic mechanisms that may not require as much water as previously believed, which may be relevant if Venus retained some water but less than Earth. Hence in our scenario a carbonate silicate cycle is hypothesized for Venus after the magma ocean phase and well before its resurfacing period to keep CO$_{2}$ largely sequestered in crustal carbonates as on modern Earth.

We are motivated to choose 1 bar atmospheres for our epochs of interest based on geological atmospheric pressure proxies for Earth that imply an atmospheric density of $\sim$0.25--1.1bar for most of the past 4 billion years \cite{Nishizawa2007,Goldblatt2009,Som2012,Som2016,Marty2013,Avice2018}. However, some models of Nitrogen cycling imply that the atmospheric pressure could have been higher in the Archean \cite{JohnsonGoldblatt2015,Mallik2018}. Differences in N$_{2}$ of factors of a few have relatively small net impacts on climate, primarily due to small decreases/increases in the pressure broadening of CO$_{2}$ lines and partly offsetting decreases/increases in Rayleigh scattering.  Given the likely similar geochemistry and volatile histories of Venus and Earth \cite<e.g.,>[]{RingwoodAnderson1977,Lecuyer2000,Treiman2009,Chassefiere2012,Rubie2015}, a similar evolution in their early composition and pressure histories is plausible. Although whether they did indeed start out with similar volatile inventories is still an open research question \cite<e.g.>[]{Horner2009}.


Additional work demonstrates that the present day CO$_2$ and N$_2$ inventories of Earth and Venus may be similar \cite{DonahuePollack1983,Goldblatt2009} if Venus is more degassed than Earth as \citeA{DonahuePollack1983} speculated. 
Early work by \citeA{Rasool1970} and \citeA{Kasting1988} estimated nearly as much CO$_2$ was locked up in carbonate rocks on Earth as exist in the atmosphere of Venus. Since there is no carbonate-silicate cycle active on Venus today most of the CO$_2$ that would otherwise be locked up in the interior is in the atmosphere, having degassed over the past several hundred million years (at least). As well, nearly 4 times as much N$_2$ is found in Venus' present day atmosphere as in Earth's when scaled by planetary mass. Earth's internal N$_2$ budget is estimated at $\sim$ 7 $\pm$ 4 times that of the atmospheric mass \cite{JohnsonGoldblatt2015}. 

However, Argon (Ar) modeling studies \cite{Kaula1999,ORourke2015} imply that Venus is less degassed than Earth, although see \citeA{DonahuePollack1983} and \citeA{Halliday2013} for alternative explanations for the current $^{40}$Ar abundance difference between Earth \& Venus.  \citeA{Watson2007} claims that $^{40}$Ar is not a reliable indicator of degassing, although ``the interpretation of their data is controversial" according to \citeA{ORourke2015}. \citeA{Halliday2013} also mentions that Venus' radiogenic $^{40}$Ar implies Venus is less degassed than Earth, but he notes that Venus' atmospheric non-radiogenic $^{36}$Ar has concentrations roughly two orders of magnitude greater than Earth's and that ``this is hard to explain unless it (Venus) is heavily degassed with a larger inventory of primordial volatiles." In further support of a degassed Venus \citeA{Halliday2013} notes that the similar Earth and Venus budgets of C and N support a degassed Venus while helping to explain the high $^{20}$Ne and $^{36}$Ar budgets of Venus. 
One of the largest stumbling blocks in deciding whether the $^{40}$Ar modeling inference is correct is a proper measurement of K/U on Venus, which is presently highly unconstrained. For this reason modeling studies generally use Earth values. \citeA{Namiki_Solomon1998} use He to confirm the $^{40}$Ar estimates, but they require a Venus in-situ mission to make the appropriate Th and U abundance measurements to characterize the geochemistry necessary for He production. One also needs to consider He escape, an under-explored area of research given that it depends not only on the ability of He to remain charged or neutral, but also whether Venus had a past magnetic field and how present day measurements can distinguish in situ He in the upper atmosphere from solar wind deposition.


Another limiting factor in comparing Earth and Venus is the lack of good constraints on the bulk water content of Earth. Some papers refer only to surface water amounts of 0.023 wt\% of the Earth's total mass, but there are compelling arguments for possibly deep reservoirs within the lower mantle or core \cite<e.g.,>[]{Raymond2006,Schmandt2014,Genda2016,Ikoma2018}. For Venus, interior water amount estimates are mostly speculation and are restricted to planetary formation studies \cite<e.g.,>[]{Ikoma2018}. 


\section{Venus' Early Evolution \& Evidence for Water}\label{section:water}

Venus \& Earth likely received similar initial water inventories during their formation histories, as has been shown in a number of works \cite<e.g.,>[]{Raymond2006}. It is also tied to their composition in general, as discussed above in Section \ref{section:atmosphere}.

The Pioneer Venus mass spectrometer measured a very high D/H ratio of 150$\pm$30 times that of terrestrial water \cite{Donahue1997} in a trapped droplet of sulfuric acid. It is the only such published in-situ measurement. Other non in-situ measurements have been made as noted in Section \ref{sec:intro}, and work continues apace \cite<e.g.>[]{Tsang2017}.  The original Venus D/H discovery paper by \citeA{Donahue1982} was titled ``Venus was Wet: A measurement of the Ratio of Deuterium to Hydrogen." This was a tantalizing prospect, but limited by our knowledge of atmospheric escape processes \cite{DonahuePollack1983,DonahueRussell1997,Donahue1999} and the D/H of delivered materials over the aeons.

It is possible that the D/H ratio is not an indicator of large amounts of water in Venus' ancient history. \citeA{Grinspoon1993} pointed out that a short residence time for water in the present atmosphere of Venus works against the primordial ocean hypothesis. \citeA{Grinspoon1993} also noted that updated theoretical calculations at that time that implied higher deuterium escape efficiency put constraints on the D/H source water of 10-15. That would rule out source material such as meteorites, comets and dust particles with high D/H ratios \cite<e.g.,>[]{Irvine2000,Charnley2009}.
As well, measurements of D/H and Xenon isotopes in Comet 67P \cite{Altwegg2015,Marty2017} imply that Earth's ocean has a much lower contribution from cometary objects than previously thought. This would also likely rule out a large cometary contribution to the high D/H ratio measured on Venus.

A number of authors have tried to model changes in the D/H fractionation over time \cite{KastingPollack1983,Gurwell1993,Gurwell1995,Hartle1996} to put some constraints on when the water was lost. Given the lack of data from Venus it is equally difficult to constrain or move these models forward.


The possibility that the high D/H ratio implies long-lived surface water is also limited by in-situ measurements. There is some circumstantial evidence of past surface water from surface emissivity observations from the Galileo NIMS instrument \cite{Hashimoto2008} and the Venus Express VIRTIS instrument \cite{Mueller2008}. These observations may imply that the highland ``tessera" regions are mostly composed of felsic rocks, and if they are indeed granitic they would have required surface water to form \cite{CampbellTaylor1983}.
\citeA{Gilmore2015,Gilmore2017} find that at least one tessera region observed with VIRTIS (Alpha Regio) appears to be more felsic than surrounding plains. This also suggests that these older stratigraphic units \cite{Ivanov1993,Gilmore1997} are granitic crustal remnants, but recent work by \citeA{Wroblewski2019} shows that parts of the Ovda Regio highland tessera are not in fact of granitic origin. However, it is not yet possible to generalize the work of \citeA{Wroblewski2019} to the entirety of tessera.

\citeA{Nikolayeva1990} and \citeA{Shellnutt2019} analyzed surface rock measurements from Venera 8. As \citeA{Shellnutt2019} explains ``it is possible that the Venera 8 probe encountered a fragment of crust that resembles a terrestrial greenstone belt." \citeA{Zolotov1997} and \citeA{Johnson2000} have also demonstrated that signatures of water in hydrous minerals may persist on the surface of Venus for long periods even after the surface morphology has changed. This motivates an in-situ mission to Venus to search for such materials, if they exist.
\citeA{Watson2007} tried to demonstrate that the $^{40}$Ar in Earth's atmosphere is related to the hydration of the oceanic lithosphere consisting of relatively Ar-rich olivine and orthopyroxene. If the results from \citeA{Watson2007} are correct, (and there is skepticism \cite{Ballentine2007}), this would lead one to believe that the $^{40}$Ar in Venus' atmosphere today implies that water oceans could have persisted for some time.

An outstanding unsolved and understudied problem is what happened in the epoch of Venus' MO as it cooled, as this may greatly affect the long-term water inventory of the planet. The timescale of the MO crystallization could be of order a few million years ($Myr$) as for Earth \cite<e.g.,>[]{Katyal2019,Nikolaou2019} or greater than 100 $Myr$ \cite{Hamano2013,Lebrun2013}. The longevity of the MO and associated hot steam and CO$_2$ atmosphere is vital to understanding the volatile history of Venus \cite<e.g.>[]{Salvador2017}. If the MO and steam atmosphere persist too long then much of the primordial water inventory of Venus could have been lost in its very early history from a stronger solar wind \cite{Chassefiere1997,Lichtenegger2016}. 
An attractive feature of the extended MO hypothesis is that it naturally solves the problem of the lack of oxygen in the present day Venusian atmosphere. This would be accomplished by sequestering the O$_2$ left behind by H$_2$O dissociation in the 
magmatic crust and upper mantle \cite<e.g.>[]{Lebrun2013,Gillmann2009,Lichtenegger2016,Lammer2018}.
We discuss how large quantities of O$_2$ can be lost after a significant period of habitability in bullet 6 of Section \ref{sec:conclusion}, in lieu of early MO losses.
It is possible that the high D/H ratio we see today \cite{Donahue1982,Donahue1997} is a relic of the early MO period. If the MO cooled quickly, then there was an opportunity to build up a surface ocean and atmosphere as is believed to have happened in Earth's early history. The question is whether Venus' surface conditions as a result of its closer proximity to the Sun would prevent the condensation of water on its surface or not. The answer is more complicated than it may seem since water can condense under hot high pressure multi-bar atmospheres. \citeA{MatsuiAbe1986} allow for temperatures up to 600K, while later work by \citeA{Liu2004} allow temperatures approaching 720K.

Additionally the answer may reside in the planet's rotation history, what role clouds played, and the outgassing rates of H$_2$O and CO$_2$. As we will show in Section \ref{sec:evolution} it is possible for Venus to reach a tidally locked state in less than a few hundred $Myr$ using constant phase lag dissipation theory, suggesting that the planet's rotation rate could have been slow early on. As shown in previous work \cite{Way2016,Way2018} as long as a planet is in the slowly rotating regime (length of day greater than $\sim$16 Earth sidereal days) its climate dynamics work to allow liquid water to persist on the surface for insolations up to $\sim$ 2.6 times that of present day Earth. This is due a large contiguous dayside cloud deck that significantly increases the planetary albedo as discussed in Section \ref{section:surface-history}.

The timing of the MO termination is critical in more than one way. If the steam and CO$_2$ atmosphere cooled sufficiently for MO crystallization to occur by the time of the Late Veneer (also referred to as ``Late Accretion") then even if Venus lost most/all of its primordial H$_2$O through escape processes \cite{Gillmann2009,Hamano2013,Lichtenegger2016} there may have been a second chance to obtain a surface ocean, albeit a shallow one. Recent work by \citeA{Greenwood2018} implies that Earth may have received as much as 30\% of its H$_2$O inventory in post-accretion impact delivery, consistent with research that shows that the entire H$_2$O budget cannot come from the late veneer \cite{MorbidelliWood2015}.
\citeA{Halliday2013} concludes that if veneers were common they should be proportional to planetary mass, and hence Venus would have received a percentage of late veneer H$_2$O similar to that of Earth.
If Venus was left dry after a long-lived magma ocean phase \cite{Hamano2013}, then this amount of H$_2$O veneer also fits within the error bounds of Venus' measured D/H ratio \cite{Donahue1982,Donahue1997}.
It should be noted that the work of \citeA{Greenwood2018} can also fit within the Ruthenium studies of \citeA{FischerGodde2017}. For a contrary point of view see \citeA{Gillmann2019}, who claim that most of the late veneer impactors would have been Enstatite/ordinary chondrites which are water-poor, as opposed to water-rich carbonaceous chondrites that would have been a mere 0-2\% of the total chondrite delivery. These contrary points of view come about because different geochemical measurements give different answers as pointed out in a number of recent works \cite<e.g.>[]{Albarede2009,FischerGodde2017,Dauphas2017,McCubbin2019,Zahnle2020}.
There are two other important caveats to consider regarding the late veneer. First the water content depends upon the composition of the accreting bodies. For example, if the late veneer was made up of a few large bodies then the variations could have been greater than if it was due to a large collection of smaller bodies. Second, the definition of the late veneer is important since it is typically associated with the accretion of bodies after the last giant impact. \citeA{Jacobson2017} has suggested that the last giant impact on Venus could have been much earlier than on Earth and this obviously affects the composition of objects making up the late veneer.

At the same time, work by \citeA{Gillmann2009}, \citeA{Morbidelli2000}, and \citeA{Raymond2006} shows that Venus' initial water inventory at formation could be as much as two terrestrial ocean's worth while large planetary embryos could deliver much more within 200 $Myr$ of formation.
If true, and if the magma ocean lifetime on Venus was shorter rather than longer, then our estimates of the water content on Venus from Pioneer Venus D/H ratios \cite{Donahue1982,Donahue1997} should be more toward the higher end, $\sim$ 16\% of a present day Earth's ocean \cite{Donahue1997}. 
However, it is not clear whether Venus' primordial water content can readily be constrained by the D/H ratio \cite{Grinspoon1987,Grinspoon1993} due to a lack of knowledge of sources and sinks over the lifetime of the planet.

Another hypothesis \cite<e.g.,>[]{RampinoCaldeira1994} states that because of its proximity to the Sun, Venus could never condense water on its surface and hence its surface temperature has always been 300K or higher \cite<see Figure 1 in>[]{RampinoCaldeira1994} and that most of this water was lost by photodissociation \cite{GoodyWalker1972}. At the same time the lack of water prevents silicate rock weathering (on Earth this removes CO$_2$ from the atmosphere), hence the CO$_2$ builds up in the atmosphere driving temperatures ever higher due to the greenhouse effect as seen today.


\section{Surface History, Impactors and Climate Evolution}\label{section:surface-history}

Understanding the surface history of Venus is crucial to constraining any theory of its long-term climate evolution. \citeA{Smrekar2018} reviews the literature on Venus' internal structure and dynamics. In this section we mostly focus on implications for the surface features we see today and how those might be consistent with a hypothesis for the long-term habitability of Venus and a transition to a more recent ($\sim$1Gyr) hothouse state.

Up to 80\% of the Venus surface has volcanic plains and tectonic structures emplaced over a relatively short geological interval as determined from crater counts \cite{IvanovHead2013,IvanovHead2015b}.  The cratering record seen in the plains regions imply surface ages ranging, for example, from $\sim$180Ma \cite{Bottke2016}, to $\sim$300Ma \cite{Strom1994} to $\sim$750Ma \cite{McKinnon1997}. The relative youth of most of Venus' surface may be the result of a large scale lithospheric overturn known as the Global Resurfacing Event (GRE),
or it may be due to the latest GRE in a long sequence of episodic resurfacing events \cite<e.g.,>{Turcotte1993,Strom1994}. For example, \citeA{Kaula1999} constructed a simple model with outgassing events staggered at time periods of 4.1, 3.8, 3.5, 3.1, 2.6, 2.1, 1.5 and 0.7 Ga constrained by $^{40}$Ar measurements. The other hypothesis for the young surface of Venus is from continuous volcanic resurfacing 
\cite<e.g.>[]{Basilevsky1997,Bjonnes2012,King2018}.

The highland tesserae may be one of the keys to understanding this history. They are of particular interest because they may contain information about past crustal differentiation and other processes prior to the loss of any surface water.  Some crater age estimates from the Magellan Mission imply that the tesserae are $\sim$40\% older than the plains \cite{Ivanov1993,Gilmore1997}. However, \citeA{Strom1994} did not agree with this conclusion.  Additional work by \citeA{Hansen2010} points to the possibility that the Ribbon Tessera Terrain are older than the surface units identified with the GRE. Later analysis by \citeA{IvanovHead2013} implied that tessera are the oldest stratigraphic unit and that they were created near the beginning of Venus' surface observable history during the ``tectonically dominated regime." 


However, {\emph how} the large basaltic plains were emplaced remains controversial. A number of authors \cite<e.g.,>[]{Herrick1994,Strom1994,BasilevskyHead1996} postulated a nearly global ($\sim$ 80\%) geologically instantaneous (10-100$Myr$) thick ($>$1km) deposition of basaltic material from volcanic type outflows (GIBVO) that would have buried older craters we cannot observe today (akin to the GRE mentioned above). The outflow depth requirements are determined by the size of the largest impact craters that would have to be completely covered. However, as \citeA{IvanovHead2013} point out it is possible that the cratering record previous to GIBVO could have also been erased in some manner. The GIBVO model was later augmented and became known as the global stratigraphy hypothesis \cite<e.g.,>[]{BasilevskyHead1996,Basilevsky1997,BasilevskyHead1998,HeadBasilevsky1998}. 
Yet another hypothesis to explain the Venus surface record was initially put forward by \citeA{Phillips1992} and is termed the Equilibrium Resurfacing Model (ERM). In this model the number of craters observed on Venus today is the result of an equilibrium between constant crater formation (via impacts) and the removal of such craters via on-going tectonic or volcanic methods. Monte Carlo calculations by \citeA{Bullock1993} and \citeA{Strom1994} demonstrated why the ERM was not feasible. \citeA{Strom1994} decided that the GIBVO was a better fit to their data, while \citeA{Bullock1993} preferred a longer timeline of 550$Myr$. More recent Monte Carlo calculations by \citeA{Bjonnes2012} show that the ERM is able to fit the observations.


\citeA{HansenYoung2007} strove to demonstrate why none of these hypotheses fit all available observational constraints. \citeA{HansenYoung2007} then proposed what they termed the Spatially Isolated Time-Transgressive Equilibrium Resurfacing (SPITTER) hypothesis to explain more of the observational constraints. It is not clear that the Venus geological community has settled on any of these hypotheses. Perhaps one of the largest problems with the global lava hypothesis is the timescale, volume and depth of the basaltic flows required, none of which have been observed on any present or previously active volcanic body in the solar system (including in Earth's past). The largest known outflow to date in Earth's history is the mid-Cretaceous Superplume \cite{Larson1991}, which is small by comparison to those envisioned to describe Venus' resurfacing. At the same time the superplume hypothesis for Venus is compelling as large amounts of CO$_2$ could have been released at the same time as the plume event \cite{CaldeiraRampino1991}. Large overturn events have been proposed as an explanation for Venus' present surface state, but in such a scenario it is possible to sequester large amounts of CO$_2$ in fresh flood basalt outflows due to enhanced planetary weatherability \cite<e.g.,>[]{Godderis2003,Cox2016}. Large Igneous Provinces (LIPs), on the other hand, can release copious amounts of CO$_2$ sequestered in some sedimentary materials \cite<e.g.,>[]{Ganino2009} while avoiding the sequestration issues of a large overturn event. LIPs have been proposed as an explanation for Venus' present day state as we will discuss below.

Previous simulations by \citeA{Way2016} showed that Venus could have had temperate conditions for nearly 2 billion years providing it had a shallow ocean of 310m in depth, slow rotation rate, and modern orbital elements. Venus might even have experienced more stable conditions than Earth in its early history since studies by \citeA{Correia_Laskar2001} and \citeA{J_Barnes2016} have shown that low obliquity states (like that of modern Venus) may be stable over billions of years and we know that the much shorter Milankovich cycles have had a strong influence on Earth's climate through time. \citeA{Deitrick2018} reviews the influence of such cycles on the climate of Earth and possible influences on exoplanets.
In addition, \citeA{Weller2018} has also shown from geological models that early Venus could have avoided glaciations more easily than early Earth, which experienced several partial or total snowball periods in its history. If long-term stable surface conditions are a requirement for life, Venus might have been more stable and allowed primitive life to fill more ecological niches more quickly than on Earth. This gives rise to the possibility that life may still exist in Venus' upper atmosphere \cite{Limaye2017}.

\citeA{Ernst2017} speculate that ``On Venus, voluminous LIP volcanism produced high levels of CO$_2$ that led to run-away greenhouse effect, and high levels of SO$_2$ that caused acid rain," but with little supporting evidence. 
\citeA{BullockGrinspoon2001} present a similar hypothesis that involves outgassing of SO$_2$ and H$_2$O that eventually drive the planet, over 100s of $Myr$, into a runaway greenhouse state, but do not mention CO$_2$. If Venus had LIP volcanism then CO$_2$ as well as SO$_2$ can be outgassed if trapped in sediments in the crust as is seen on Earth \cite<e.g.,>{HeadCoffin1997,Hansen2007,Ernst2017,ErnstYoubi2017}. Hence if Venus had an earlier epoch of liquid water habitability then it is logical to assume that CO$_2$ would have been trapped in the crust of the planet in the same way it is trapped on Earth today and LIP volcanism would have been the means to release that CO$_2$ into the atmosphere.

However, as noted in \citeA{MacDonald_Wordsworth2017} when the surface temperature is warmer (T$>$300K, see their Figure 2) more water vapor is injected into the stratosphere, which stabilizes the lapse rate. Such warm climates (as seen in the Venus models herein) would prevent the largest plumes from injecting SO$_2$ into the stratosphere, allowing CO$_2$ warming without offsetting cooling by H$_2$SO$_4$ aerosols.

Another well known mechanism to get Venus from a cool clement state to its present day hot and dry state was proposed by a number of authors  \cite<e.g.>[]{Ingersoll1969,KastingPollack1983,Kasting1984,Kasting1988,TaylorGrinspoon2009} who speculated that water loss via upper atmospheric dissociation and then hydrogen escape would have eventually made the planet dry. Then, as stated in \citeA{TaylorGrinspoon2009} ``With the loss of water, the removal mechanism for CO$_2$ would be eliminated, and carbonate rocks on the surface would presumably eventually be subducted and lost to thermal decomposition, with the CO$_2$ being irreversibly returned to the atmosphere through outgassing."
This model fits in with more recent research by \citeA{Wordsworth2016} who states that the oxygen left over would eventually find its way to oxidize the mantle and change its redox state, allowing for enhanced nitrogen outgassing which is compatible with the nearly 3 bars of N$_2$ we see in Venus' atmosphere today (also see review by \citeA{Lammer2018}).  However, an alternative hypothesis is proposed by \citeA{Gillmann2009}, who suggest that the oxidation of the mantle occurred in the first 100 $Myr$ of Venus' history. They assume the surface was never cool enough to allow liquid water to condense. The water would again be photodissociated and the hydrogen would have been lost to space \cite{Lichtenegger2016}. The leftover oxygen would have dissolved in the magma ocean.

\citeA{Genda2005} have proposed that the lack of water on Venus and in Venus' protoplanetary impactors in its early history (in contrast to that of Earth and its water rich impactors) would explain differences in most of the noble gas abundances between Venus and Earth because oceanic protoplanets would enhance atmospheric loss, implying that Venus' original noble gas abundant proto-atmosphere survived to present day on Venus, unlike that of Earth. A lack of water being detrimental to subductive plate tectonics (see Section \ref{sec:conclusion}).
\citeA{Sakuraba2019} have also attempted to get the presently observed nitrogen and noble gas abundances via impact degassing and atmospheric erosion \cite<also see work by,>[]{Pham2011}, but unlike \citeA{Genda2005} they believe late accretion may have further influenced the atmosphere of Venus.

More recent work  by \citeA{Gillmann2016} show that large impactors (400-800km in diameter) can cause atmospheric erosion and escape and deposit energy in the crust and mantle. They believe the latter can
cause a thermal anomaly in the crust and mantle triggering large scale volcanic events at the impact region and the antipode. This in turn may deplete the upper mantle of volatiles and lead to water loss in the early atmosphere, or conversely provide a volatile heavy atmosphere with extreme temperatures for billions of years. In a sense this is similar to a theory by \citeA{Davies2008} who propose a mega-collision (akin to that of the Earth's moon-forming impact) to dry out the interior of the planet. But thus far no large Venus impactor simulations have been utilized to examine such a scenario, as has been done for Earth's moon-forming collision \cite<e.g.,>[]{Canup2004}.

To summarize, a number of mechanisms exist by which early Venus could have condensed liquid water on its surface. The key ingredient is that it must have been cool enough for long enough in its early history. As shown by \citeA{Yang2014,Way2016,Way2018}, the rotation rate of a planet greatly affects its climate dynamics. Specifically, for very slow rotation a large contiguous water cloud forms at the substellar point, increasing the Bond albedo markedly and keeping surface temperatures moderate for insolation values up to nearly three times that of modern Earth's 1361 W m$^{-2}$.  In Paper I \citeA{Way2016} we demonstrated that early Venus could have had consistently habitable conditions throughout its early history if it began with sufficiently slow rotation.  In the next section we review what is understood about the possible evolution of Venus' spin-orbit state.

In our scenario, early Venus' has the earliest consistent liquid water habitability in the solar system followed by Earth and then Mars. This is a broader statement of the Faint Young Sun Paradox (FYSP), the challenge of explaining how early Earth, not to mention Mars, could have been warm and wet early in their histories when the Sun was 25-30\% dimmer than today \cite<e.g.,>{Feulner2012}. There is still debate in the ancient Earth GCM community about the actual composition and thus temperature of early Earth's atmosphere given observational proxies for CO$_2$ that span orders of magnitude, though models suggest that the range encompasses several viable scenarios \cite<e.g.,>[]{Charnay2013,Wolf_Toon2013,Kunze2014,LeHir2014,Charnay2017,Krissansen-Totton2018}. 
These GCM studies and most proxies \cite<e.g.,>[]{Spencer2019} are from the Archean rather than the late Hadean, but there is some evidence that habitable surface conditions existed well back into the Hadean \cite<e.g.,>[]{Harrison2009,Arndt2012}.

It is interesting to note that recent atmospheric pressure proxies from the late Archean imply an atmospheric pressure less than half that of today \cite{Som2012,Som2016}. Atmospheres thinner than modern Earth's are less likely to avoid snowball conditions, yet the literature above notes that there is geological evidence that Earth was not in a snowball state during much of the late Archean that the pressure proxies correspond to. Regardless, for this reason we feel is it necessary to explore the possibilities of lower atmospheric surface pressures in Venus' climatic history as described for Simulations 26-30 in Section \ref{sec:methods}.

The FYSP for Mars remains difficult to resolve \cite<e.g.,>[]{Wordsworth2016AREPS} partly due to the fact that 3-D GCMs have traditionally struggled to consistently sustain large-area liquid water conditions over millions of years \cite<e.g.,>[]{Goldblatt2009,Kasting2010,Kienert2012,Feulner2012,Haqq-Misra2008} without snowball type conditions.
Long-standing solutions involving large amounts of atmospheric CO$_2$ are inconsistent with unobserved carbonate deposits expected from such CO$_2$ dominated atmospheres \cite{Shaw2018} and are insufficient in isolation to produce above-freezing conditions. One possible solution to the lack iof surface carbonates was proposed by \citeA{Kasting2012}. Other solutions to Mars' FYSP exist that involve H$_{2}$ with CO$_{2}$ as the background gas \cite<e.g.>{Wordsworth2017,Ramirez2014b,Ramirez2018,Haberle2019}, although presently there appears to be little consensus in the community.

\section{Rotation and Obliquity evolution}\label{sec:evolution}

To the best of our abilities we would like to constrain the obliquity and rotational history of Venus to better constrain these important inputs for climate models. This is limited by the absence of any direct information about Venus' initial rotation and obliquity and the fact that impacts likely play a significant role in the early rotational history of the terrestrial planets \cite<e.g.,>[]{LissauerKary1991,DonesTremaine1993}. On Earth a variety of means exist to obtain some constraints using dynamical modeling combined with geological data when available \cite<e.g.,>[]{Hays1976,Park1987,Imbrie1992,Matthews1997,Petit1999,Palike2000,Palike2004,Olsen2019} and there has been modest success doing the same for Mars \cite<e.g.,>[]{Cutts1982,Laskar2002Mars,Laskar2004,Byrne2009,Dickson2015,Bierson2016}. For Earth, an additional constraint is provided by the Moon, which has predictably affected the evolution of Earth's rotation and damped obliquity excursions over its history \cite{ZW1987,Lissauer2011}.  However, until and unless geological observables become available to constrain dynamical models, only plausible scenarios for the rotational and obliquity history of Venus can be defined.

\citeA{vanHoolst2015} summarizes much of the literature on the rotational evolution of Venus throughout its history. We summarize some of the work on this subject below and add some additional estimates.  First we look at the history of studies of the possible spin evolution of Venus.

In the 1960-70s several authors investigated the possibility that Venus' rotation period was correlated with its synodic period \cite{Goldreich1966,Gold_Soter1969,Gold_Soter1979}. \citeA{Goldreich1966} states, ``the presence of the Earth may have stabilized the sidereal rotation period of Venus at the value of 243.16 days retrograde." An equilibrium between the atmospheric and body tide of Venus was first proposed by \citeA{Gold_Soter1969} to explain Venus' non-synchronous rotation period, based on the incorrect belief at that time that Venus always showed the same face at each inferior conjunction with Earth as proposed by \citeA{Goldreich1966}.

The first work to analytically look at Venus' rotation rate and the role of atmospheric tides was by \citeA{Ingersoll1978} who extended the earlier work of Lord Kelvin \citeA{Thompson1882}, \citeA{ChapmanLindzen1970} and \citeA{Munk_MacDonald1960}. They mention that ``Venus probably originated with a retrograde rotation in order to have evolved to the current retrograde state." In the 1980s this work was further extended in a series of papers  \cite{Dobrovolskis1980a,Dobrovolskis1980b,Dobrovolskis1983}.
It was clear that Venus' rotation rate was probably determined by a balance between the solid body tidal dissipation and the thermal tides of its thick atmosphere with the sun. Core-mantle friction (CMF) can also play an important role in slowing the spin rate of Venus, as first explored by \citeA{Goldreich_Peale1970}.
\citeA{Goldreich_Peale1970} were also the first to demonstrate that core-mantle viscous coupling can drive the obliquity to 0$^\circ$ when less than 90$^\circ$ and to 180$^\circ$ if it is greater than 90$^\circ$ over time.


This remained the state of understanding of Venus' rotational history until the early 2000s when the long-term evolution of its spin state of Venus was investigated in a series of papers by \citeA{Correia_Laskar2001,Correia2003a,Correia2003b}, who suggested
that Venus may have rotated faster in the past, and possibly prograde.
It also became clear that at faster spin rates CMF plays an important role in slowing the rotation of the planet, but less so at slower spin rates.

Once a planet is spinning more slowly CMF may play an important role in obliquity variations \cite<e.g.,>[]{Correia2003a}.
\citeA{Correia_Laskar2001} explored a number of stable obliquity and spin states of Venus while more recent work by \citeA{J_Barnes2016} has investigated how stable the obliquity of Venus might be though time.

The work on the thermal tides of Venus had led researchers to assume that its effects would be minor (as it is for Earth) for atmospheres of modest density (e.g. 1 bar). However, more recent work by \citeA{Leconte2015} has demonstrated that thermal tides arising from even 1 bar atmospheres can be significant depending on the distance to the host star and the host star's mass. \citeA{Leconte2015} show that even if modern Venus had a 1 bar atmosphere the tidal torques would still be quite significant.

\begin{figure}[ht!] \centering
\includegraphics[width=\textwidth]{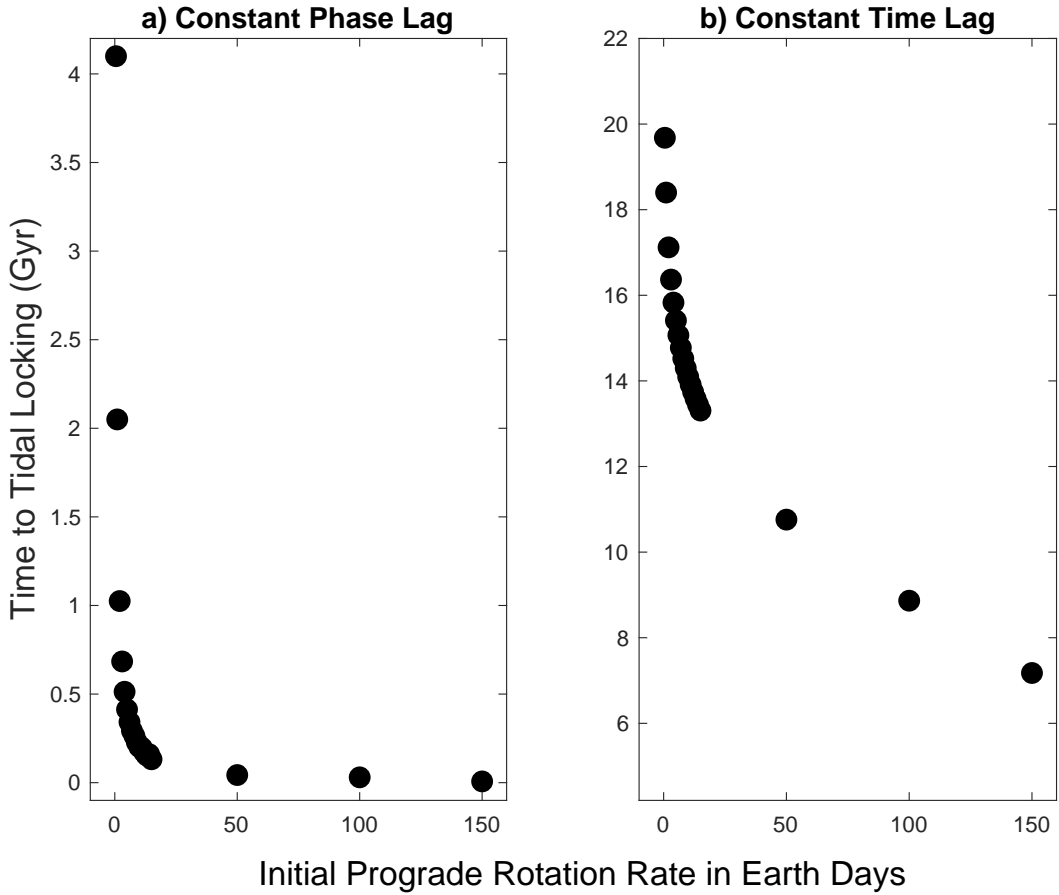} 
\caption{\small Venus tidal locking timescales using equilibrium tide constant phase (a) and time (b) lag theory. Note that the y-axes have different limits.} \label{fig1}
\end{figure}

\citeA{Barnes2017} used an equilibrium tide model with a constant phase lag (CPL)  to find that Earth could have ended up tidally locked today (after 4.5Gyr) had it started with a rotation rate of 3 Earth Days or longer (the latter more likely if Earth had no satellite). We have applied the same Equilibrium Tide Model (EqTide https://github.com/RoryBarnes/EqTide) from \citeA{Barnes2017} to Venus to explore how long it would take Venus to reach a tidally locked state only from solid body tides. As shown in Figure \ref{fig1} using CPL theory we find that Venus could have been tidally locked within 684$Myr$ if it started with a prograde rotation period of 3 Earth days and zero obliquity.
Unfortunately the EqTide model we utilize does not support retrograde spin states, but we expect the differences to be minor. We will continue to explore these issues in a future work using the simulator vplanet \cite{Barnes2019} once this functionality is added. Figure \ref{fig1} gives further examples for CPL and Constant Time Lag (CTL) theory results using EqTide. For input parameters we assume that the tidal dissipation factor Q=12 and Love number of degree 2 k$_2$=0.3. These are the same numbers used for the modern Earth in \citeA{Barnes2017}.
Recent work by \citeA{Henning2014} demonstrates that our choice for Q may not be unreasonable for Venus. \citeA{Henning2014} give estimates of Q for Earth-like planets (see their Fig 15, top-center-row plot) with orbital periods from 0 to 200 days. Venus' 224 d period is slightly outside the range they explore (but can be anticipated from the trend visible in their figure). Our assumption of Q=12 is not far off the \citeA{Henning2014} `Warm Earth 2' estimate in their Fig 15. As an aside, Q and k$_2$ are poorly constrained for present-day Venus. We have even fewer constraints on these values for an ancient Venus, but perhaps those values would be more Earth-like than present day Venus. For example, present day Venus' time lag may not be the same as Earth's because of higher internal temperatures \cite{MacDonald1962,Henning2014}.
Historically \citeA{Goldreich_Soter1966} estimated that Q$<$17 for Venus, \citeA{Lago1979} had values up to Q$\sim$40 while \citeA{Leconte2018} estimate Q$\sim$100. 

More recently Venus' tidal love number was estimated by \citeA{Konopliv1996} using Magellan and Pioneer Venus Orbiter data to be k$_2$=0.295$\pm$0.066 implying the core is liquid \cite{Yoder1997}.
Work by \citeA{Zhang1992,Xia2002} have estimated k$_2$=0.18 $\sim$ 0.26. A smaller value (k$_2$=0.17) would imply a solidified iron core which is not consistent with \citeA{Konopliv1996}. Modeling work by \citeA{Dumoulin2017} 
are consistent with the work of \citeA{Konopliv1996} as well as our own modeling choices (discussed above) of Q=12 and Love number of degree 2 k$_2$=0.299 (see Table 3 in \citeA{Dumoulin2017}).
Regardless, if one uses higher values of Q and/or lower values of k$_2$ for ancient Venus it is sufficient to say that equilibrium tide theory predicts that the CPL and CTL for Venus estimates for tidal locking will be longer than those presented in Figure \ref{fig1}. The values in Figure \ref{fig1} then represent \emph{lower limits} to tidal locking for a given starting rotational period.
As a caveat there is a debate in the dynamics community about the appropriateness of the CPL and CTL approaches \cite{Efroimsky2009,Efroimsky2013,Touma_Wisdom1994,Greenberg2009}. so these tidal locking timescales should be viewed with some caution in the context of
the CPL and CTL models used herein. Of course we do not take into account magnetic braking to see how the Sun's natural spin-down might affect the tidal evolution of Venus, nor do we assume that Venus' orbital characteristics would have changed over the timescale of our calculations, the latter being one of the criticism when applying CPL/CTL to evolving systems \cite{Efroimsky2009,Efroimsky2013}.

Recent work \cite{Green2019} has investigated the influence of a hypothetical shallow ocean on Venus (water equivalent layers of $\sim$330 meters deep and 830m) using present day topography and a range of initial rotation periods.
The most dissipative scenario predicts a slow down of 72 days per million years.
The latter result may be surprising until one recalls that tidal dissipation in Earth's oceans is larger than that of the Earth's solid body tides \cite<e.g.,>[]{Munk_MacDonald1960}.


It is likely that Venus was initially a prograde spinning body like the other 3 terrestrial planets in our solar system. The prograde hypothesis goes back at least to \cite{Alfven1964} and more recent work by \citeA{LissauerKary1991,DonesTremaine1993,KaryLissauer1995} would also support the idea of a primordial prograde Venus, barring the effects of a late large impactor as discussed in those works and that below. From that starting point we find the following perhaps the most compelling answer to Venus' present day spin state. As shown above there are models that can drive the planet toward a tidally locked state rather quickly. Core-mantle friction damps
obliquity perturbations which drive the spin rate to sub-synchronous prograde rotations. Then atmospheric tides would reverse the spin to a retrograde equilibrium.
These atmospheric tides continue to prevent the planet from being tidally locked and that is the state the planet has been in since that time as shown in some of the work of \citeA{Correia_Laskar2001}.

Yet it has long been speculated that Venus' current retrograde rotation state is the result of a large impactor early it its history \cite{McCord1968,Singer1970,FrenchSinger1971,Counselman1973,Burns1973,WardReid1973,Harris1978,Alemi2006,Davies2008}. The large impactor hypothesis may also explain a possibly very dry Venus interior, a lack of oxygen in the atmosphere and reconcile the $^{40}$Ar results that imply it is less degassed than Earth \cite{Davies2008}. However, as pointed out in \citeA{WardReid1973} it is possible for a small impactor (less than 1\% of the Moon's mass) to drive Venus retrograde if the planet has already spun down considerably due to tidal dissipation with the Sun (Figure \ref{fig1}).

Unfortunately, there is little hope that we will ever truly know the rotation rate of Venus through time without a way to either measure its ``Geological Orrery" as on Earth \cite{Olsen2019} and tie that to dynamical models, or find evidence that an impactor played a role in its rotational and hence geochemical evolution as seen in investigations on Earth related to late accretion and its effect on different isotope abundances \cite<e.g.,>[]{Varas-Reus2019}. In fact \citeA{Brasser2016} and \citeA{Mojzsis2019} prefer the hypothesis that the Earth's Late Veneer was mainly delivered by a single Pluto- or Ceres-sized impactor. Hence if a larger object was involved in the late evolution of Venus' spin or obliquity it may be possible to detect its geochemical fingerprints in a future in-situ mission.


For \emph{most} of our climate simulations we assume, rightly or wrongly as discussed above, that Venus has had the same retrograde rotation and its present day obliquity for the past 4.2$\times$10$^9$ years. We have included a few faster rotation rates that approach the `fast rotators' described in \citeA{Yang2014,Way2018}, meaning sidereal day lengths of 16 and 64 times that of present day Earth.

\section{Methods}\label{sec:methods}

All our simulations use \rocke{} \cite{Way2017} a three-dimensional (3-D) General Circulation Model (GCM) developed at the NASA Goddard Institute for Space Studies (GISS). Radiative transfer in ROCKE-3D uses k-coefficients derived from the HITRAN 2012 line list, as well as the MT-CKD 3.0 water vapor continuum and CO2 collisionally-induced absorption and sub-Lorentzian line shapes, as described in references cited in \cite{DG2019}.  HITRAN 2012 is accurate for temperatures below 350 K, as shown by \cite{Kopparapu2013}.  As we will discuss below, a small number of our simulations exhibit a growing radiation imbalance with time, indicative of continually rising temperatures and a transition to a runaway greenhouse.  If our objective were to determine the threshold for a runaway, we would need to use the more comprehensive HITEMP line list for these simulations, as discussed by \cite{Kopparapu2013}.  Our purpose, though, is simply to identify such cases and exclude them from further analysis. 

Most simulations use modern Venus' current orbital parameters, slow retrograde rotation period (-243 Earth sidereal days in length) and orbital period (224 sidereal days). 
In Paper I (plotted herein with ID = B) we looked at a faster rotation period
(16 x modern Earth's sidereal day length) to see how the planet's early climate might have responded, and we also look at 16 \& 64 day retrograde rotation periods in Venus' early history in this work. Our focus is on changing insolation, topography, land/sea mask, surface water availability and atmospheric constituents. We motivate our choices below.

All simulations discussed below are outlined in Table
\ref{table:Experiments}. Our focus is on the retention and stability of surface liquid water on Venus over time.  Therefore we simulate four types of planets with surfaces that differ in the amount of water they contain and how that water is allowed to interact with the atmosphere. \rocke{} allows for 3 types of surface water: Soil moisture at and beneath the surface with no standing bodies of water; ``dynamic" lakes whose depth and area vary with time and that can appear or disappear as the competition between precipitation and evaporation dictates; and deeper oceans with permanent boundaries and an effectively infinite source of water for the atmosphere.  All planets with oceans are fully dynamic. For more details on the capabilities of such oceans see \citeA{Way2017}.
\begin{itemize}
    \item Arid Venus: This planet has modern Venus topography, but only contains 20cm of water in the subsurface soil layers, soil consisting of 100\% sand, and no surface standing water at the start of the simulation. The atmosphere is initialized with zero water vapor and an isothermal temperature profile at 300K. This initial condition is similar to that of \citeA{Kodama2019,Abe2011} who attempt to limit the amount of water vapor in the atmosphere (a strong greenhouse gas) and subsequently push the inner edge of the habitable zone farther inward. However, \citeA{Kodama2019,Abe2011} use modern Earth's rotation rate for all their experiments.
    
    \item 10m-Venus: Uses modern Venus topography and places a  10 meter liquid water-equivalent layer in the lowest lying topographic areas. These are treated by the model as lakes, which have no circulation. The soil is a 50/50 sand/clay mix as used in \citeA{Yang2014,Way2016,Way2018}.
    
    \item 310m-Venus: Similar to 10m-Venus, except with a 310 meter water equivalent layer again spread in the lowest lying regions. This is the same topography used in \citeA{Way2016} simulations A,B and D.
    
    \item 158m-Aqua: This is a simple aquaplanet configuration that is commonly used in the exoplanet community. It uses a fixed 158 meter deep ocean, which corresponds to the bottom of the fifth layer of the \rocke{} ocean model. It is a bit shallower than the mean depth of the 310m-Venus ocean, and therefore comes into equilibrium a bit faster while still having a similar heat capacity, while including horizontal heat transport as well as wind-driven and thermohaline overturning circulations.
    
    \item 310m-Earth: Similar to 310m-Venus, but using a modern Earth-like land/sea mask with a 310m deep bathtub dynamic ocean (i.e., every ocean grid cell is of a fixed depth of 310m). We call this an Earth-like land/sea mask since it is not exactly modern Earth, but has some modest changes as shown in \citeA{Way2018} Figure 8.
    
\end{itemize}

The five planets above are then given four types of atmospheres and four different insolations as described below:

\begin{itemize}
\item Simulations 1-5: These have a 10 bar 100\% CO$_2$ atmosphere using a solar spectrum and insolation from 4.2Ga from the work of \citeA{Claire2012}. CO$_2$ was probably the dominant gas in Earth's early atmospheric evolution \cite<e.g.,>[]{Kasting1993a}. We pick atmospheric pressures
of 1 bar (see next bullet point) and 10 bar to cover the \citeA{Kasting1993a} ranges (see their Fig 2). Our 10 bar results either equilibrate at a temperature beyond that at which our radiative transfer is accurate, or do not reach equilibrium and the temperatures attained at the time the experiments were terminated are already beyond the upper limits of our radiation tables. We report the results of these experiments in Table 2 below simply as a guide for future research, but we exclude them from our analysis in Figs. 2-8. The 10 bar simulations use a modern Venus rotation rate and obliquity.

\item Simulations 6-10: Similar to Simulations 1-5, but these use a 1 bar  97\% CO$_2$ and 3\% N$_2$ atmosphere at 4.2Ga. 

\item Simulations 11-15: As in Simulations 6-10 but with a rotation period of --16 sidereal Earth days to place the planet on the edge of the fast rotator regime as described in \citeA{Yang2014,Way2018}. This allows us to explore the possibility that the planet was rotating more quickly in its early history than today. The choice of a retrograde rotation rate was chosen to be consistent with the present day retrograde rotation, but unpublished simulations with prograde rotation rates with these values produce very similar temperatures. Note that the work of \citeA{Correia_Laskar2001,Correia2003b} indicate that prograde rotation rates of 16 days for Venus put its spin axis (obliquity) in a possibly chaotic regime. However, other work by \citeA{J_Barnes2016} indicate that low obliquity retrograde rotation rates generally have more stable spin axes. Even if there are spin axis variations on geological timescales, it is not possible for us to model those here given that \rocke{} simulations are limited to $<$ 10,000 years in length.

\item Simulations 16-20: As in Simulations 6-10 but with a rotation period of --64 sidereal Earth days. This allow us to explore the possibility that the planet was rotating somewhat more quickly in its early history but still in the slowly rotating dynamical regime. Again, prograde rotation rates were also used in unpublished results and have similar global surface temperature values.

\item Simulations 21-25: These simulations use an atmospheric composition and pressure very similar to modern Earth, namely an N$_2$-dominated atmosphere with 400ppmv CO$_2$ and 1ppmv CH$_4$ with a 1013mb surface pressure. They also use a solar spectrum and insolation at 2.9Ga from \citeA{Claire2012}. The rotation rate is the same as modern Venus.

\item Simulations 26-30: Similar to Simulations 21-25, but with a lower atmospheric surface pressure of 250mb. This is again in the interest of comparative climatology since the Archean atmospheric pressure proxy work
of \citeA{Som2012,Som2016} suggests that Earth may have had a surface pressure similar to 250mb at this time.

\item Simulations 31-35: Similar to Simulations 21-25, but now using a solar spectrum and insolation from 0.715Ga from the work of \citeA{Claire2012}.

\item Simulations 36-40: Again, similar to Simulations 21-25, but now using a modern solar spectrum and insolation.

\item Simulations 41-45: Similar to Simulations 21-25, but now using a modern solar spectrum, but with insolation set to 1.26 times Venus' present day insolation (2.4 times modern day Earth's insolation) to test the boundaries of the inner edge of the habitable zone as in \citeA{Way2018}. Two of these simulations are also out of radiation balance and trending toward a runaway greenhouse state, and thus we do not analyze them further.

\end{itemize}

\begin{table}[ht!]
\caption{Experiments}
\label{table:Experiments}
\scriptsize
\begin{tabular}{|l|l|l|l|l|c|c|c|c|c|} 
\hline
ID&Topography$^a$&Epoch&Insolation$^b$&P$^{c}$&Spin&N$_2$ &CO$_2$&CH$_4$& Soil\\
  &              & Ga  &S0X/W m$^{-2}$& bar   &days &ppmv  &ppmv &ppmv  & Type$^{d}$\\
\hline
01& Arid-Venus   &  4.2& 1.396/1913.6 & 10 &  -243&0    &1000000&  0 & S\\
02& 10m-Venus    &  "  & "            &  " & "    &"     &"     &  " & S/C\\
03& 310m-Venus   &  "  & "            &  " & "    &"     &"     &  " & S/C\\
04& 158m-Aqua    &  "  & "            &  " & "    &"     &"     &  " & - \\
05& 310m-Earth   &  "  & "            &  " & "    &"     &"     &  " & S/C\\
\hline
06& Arid-Venus   &  4.2& 1.396/1913.6 &  1 & -16  &43000 &970000&  0 & S\\
07& 10m-Venus    &  "  & "            &  " & "    &"     &"     &  " & S/C\\
08& 310m-Venus   &  "  & "            &  " & "    &"     &"     &  " & S/C\\
09& 158m-Aqua    &  "  & "            &  " & "    &"     &"     &  " & -\\
10& 310m-Earth   &  "  & "            &  " & "    &"     &"     &  " & S/C\\
\hline
11& Arid-Venus   &  4.2& 1.396/1913.6 &  1 & -64   &43000 &970000& 0 & S\\
12& 10m-Venus    &  "  & "            &  " & "    &"     &"     &  " & S/C\\
13& 310m-Venus   &  "  & "            &  " & "    &"     &"     &  " & S/C\\
14& 158m-Aqua    &  "  & "            &  " & "    &"     &"     &  " & -\\
15& 310m-Earth   &  "  & "            &  " & "    &"     &"     &  " & S/C\\
\hline
16& Arid-Venus   &  4.2& 1.396/1913.6 &  1 & -243 &43000 &970000&  0 & S\\
17& 10m-Venus    &  "  & "            &  " & "    &"     &"     &  " & S/C\\
18& 310m-Venus   &  "  & "            &  " & "    &"     &"     &  " & S/C\\
19& 158m-Aqua    &  "  & "            &  " & "    &"     &"     &  " & -\\
20& 310m-Earth   &  "  & "            &  " & "    &"     &"     &  " & S/C\\
\hline
21& Arid-Venus   &  2.9& 1.47/2001.0  &  1 & -243 &1012599&  400&  1 & S\\
22& 10m-Venus    &  "  & "            &  " & "    & "     &  "  &  " & S/C\\
23& 310m-Venus   &  "  & "            &  " & "    & "     &  "  &  " & S/C\\
24& 158m-Aqua    &  "  & "            &  " & "    & "     &  "  &  " & -\\
25& 310m-Earth   &  "  & "            &  " & "    & "     &  "  &  " & S/C\\
\hline\hline
26& Arid-Venus   &  2.9& 1.47/2001.0  & 0.25 & -243 &1012599&  400& 1& S\\
27& 10m-Venus    &  "  & "            &  "   & "    & "     &  "  & "& S/C\\
28& 310m-Venus   &  "  & "            &  "   & "    & "     &  "  & "& S/C\\
29& 158m-Aqua    &  "  & "            &  "   & "    & "     &  "  & "&-\\
30& 310m-Earth   &  "  & "            &  "   & "    & "     &  "  & "& S/C\\
\hline
31& Arid-Venus   &0.715& 1.71/2358.9  &  1 & -243 &1012599&  400&  1 & S\\
32& 10m-Venus    &  "  & "            &  " & "    & "     &  "  &  " & S/C\\
33& 310m-Venus   &  "  & "            &  " & "    & "     &  "  &  " & S/C\\
34& 158m-Aqua    &  "  & "            &  " & "    & "     &  "  &  " & -\\
35& 310m-Earth   &  "  & "            &  " & "    & "     &  "  &  " & S/C\\
\hline
36& Arid-Venus   &0.0  & 1.9/2601.0   &  1 & -243 &1012599&  400&  1 & S\\
37& 10m-Venus    &  "  & "            &  " & "    & "     &  "  &  " & S/C\\
38& 310m-Venus   &  "  & "            &  " & "    & "     &  "  &  " & S/C\\
39& 158m-Aqua    &  "  & "            &  " & "    & "     &  "  &  " &-\\
40& 310m-Earth   &  "  & "            &  " & "    & "     &  "  &  " & S/C\\
\hline
41& Arid-Venus  &Future& 2.4/3266.0   &  1 & -243 &1012599&  400&  1 & S\\
42& 10m-Venus    &  "  & "            &  " & "    & "     &  "  &  " & S/C\\
43& 310m-Venus   &  "  & "            &  " & "    & "     &  "  &  " & S/C\\
44& 158m-Aqua    &  "  & "            &  " & "    & "     &  "  &  " &-\\
45& 310m-Earth   &  "  & "            &  " & "    & "     &  "  &  " & S/C\\
\hline
D$^d$& 310m-Venus   & 2.9 & 1.47/2001.0  &  1 & -16  &1012599&  400&  1 & S/C\\ 
\hline
\end{tabular}
\vspace{0.1cm}
\newline

$^a$Topography: Arid-Venus=Only Ground Water, no surficial reservoirs, 20cm water in soil, with modern Venus topography; 10m-Venus=10m Water Equivalent Layer (WEL) spread in lowest elevations as lakes with modern Venus topography; 310m-Venus=310m deep ocean with modern Venus topography; 310m-Earth=Modern Earth-like topography with 310m deep ocean; 158m-Aqua=158m deep aquaplanet.\newline
$^b$Insolation: S0X = multiple of amount that Earth receives today in insolation (S0=1361 W m$^{-1}$).\newline
$^c$ Pressure in bar. $^d$ S=100\% Sand, S/C=50/50\% Sand/Clay, - = Not Applicable, 100\% ocean.\newline
$^d$ Simulation D from Paper 1 \cite{Way2016}. Most similar to ID 8.
\end{table}

Most simulations except 158m-Aqua use a fixed ground albedo of 0.2 (thermal conductivity = 0.26 W m$^{-1}$ K$^{-1}$) and 50/50 mix of sand/clay soil following the work of \citeA{Yang2014,Way2016,Way2018}. The Arid-Venus simulation uses the same albedo (0.2), but utilizes a 100\% sand soil, rather than the sand/clay mix in other simulations. The advantage of using sand is that it more quickly loses and absorbs water. This allows the ground hydrology to come into balance more quickly than other soil types. This is because in the Arid-Venus simulations we are focused on water availability to/from the atmosphere from/to the soil and hence the amount of total water vapor acting as a greenhouse gas in the atmosphere.



\section{Results and Discussion}\label{sec:results}

Simulations 1-5, all with a 10 bar pure CO$_2$ atmosphere and 4.2 Ga insolation, are uniformly uninhabitable regardless of the surface water reservoir and topography (see Table 2 in the Supplementary Information).  The driest planet (Arid Venus) does reach equilibrium, but with a surface temperature of 262$^\circ$C, well above the accuracy limits of the radiation parameterization used by \rocke.  The other four planets are also well above 100$^\circ$C at the point at which they were terminated and are not converging to equilibrium. Given the greater water reservoirs in these simulations, they are likely to be approaching a runaway greenhouse state.

\begin{figure}[ht!] \centering
\includegraphics[width=\textwidth]{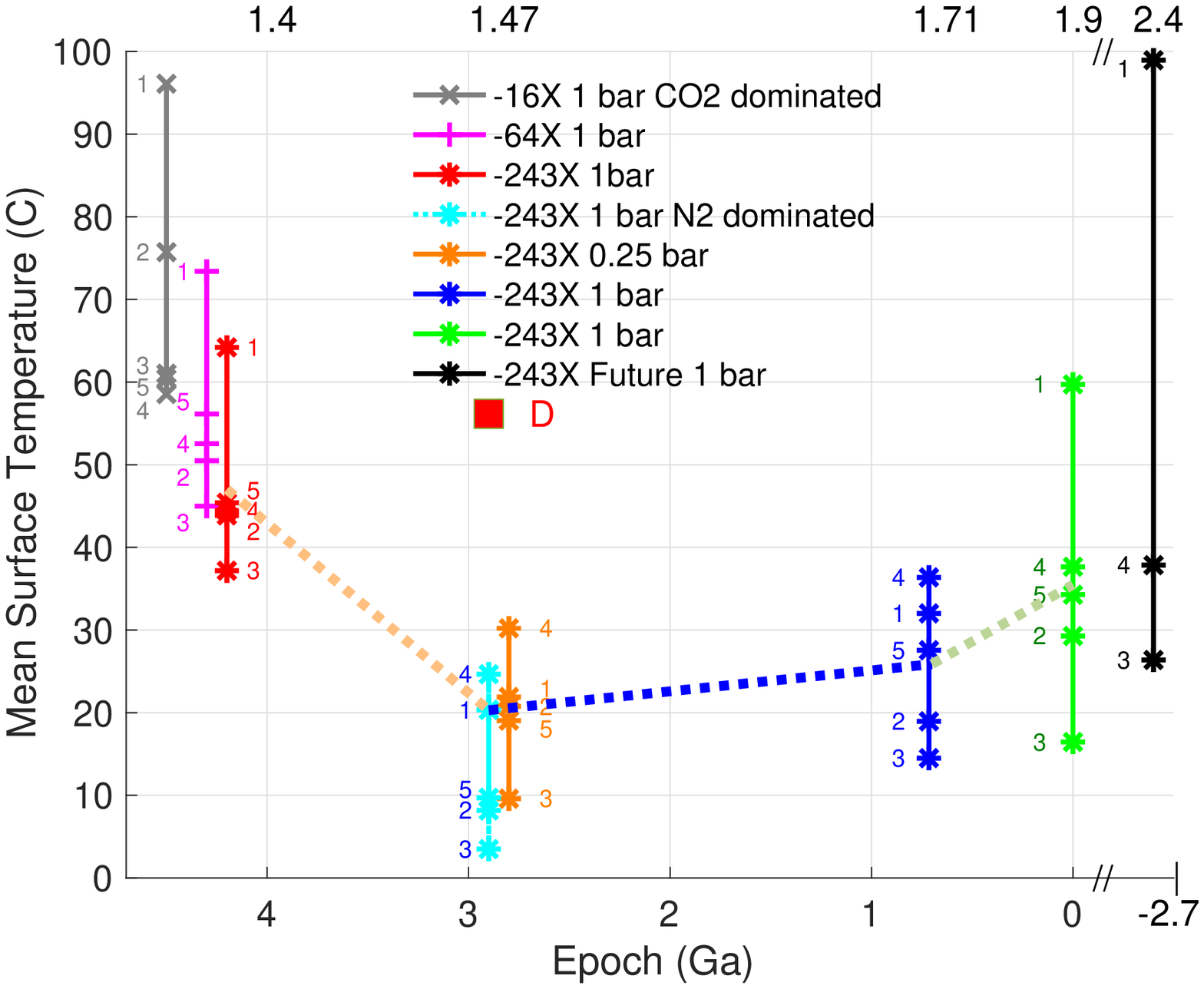} 
\caption{\small Atmospheric evolutionary scenarios for Venus. Top x-axis is insolation relative to modern Earth (1.4 = 1.4 $\times$ 1361 W m$^{-2}$). Note that the gray, magenta and red data in the left-most section of this plot are all for 4.2 Ga simulations. They are separated purely for visual effect and do not reflect differences in epoch or insolation. The same is true for the cyan, and orange data, all of which correspond to 2.9 Ga.  The numbers oriented vertically along each set of simulations correspond to the different water reservoir/topography types: 1=Arid-Venus, 2=10m-Venus, 3=310m-Venus 4=158m-Aqua, 5=310m-Earth. See Table \ref{table:results}.} \label{fig2}
\end{figure}

In Figure \ref{fig2} we show several different possible evolutionary scenarios for Venus derived from the other experiments in Table \ref{table:Experiments}. In all such scenarios we assume that Venus had surface liquid water in varying amounts at model start, as described in Section \ref{sec:methods}.  The colors in this figure differentiate groups of simulations with different insolation, rotation, surface pressure, and/or atmospheric compositions, while the numbers 1-5 and corresponding symbols for each color delineate the range of climates obtained for different surface water reservoir and topography assumptions.

\begin{figure}[ht!] \centering
\includegraphics[width=\textwidth]{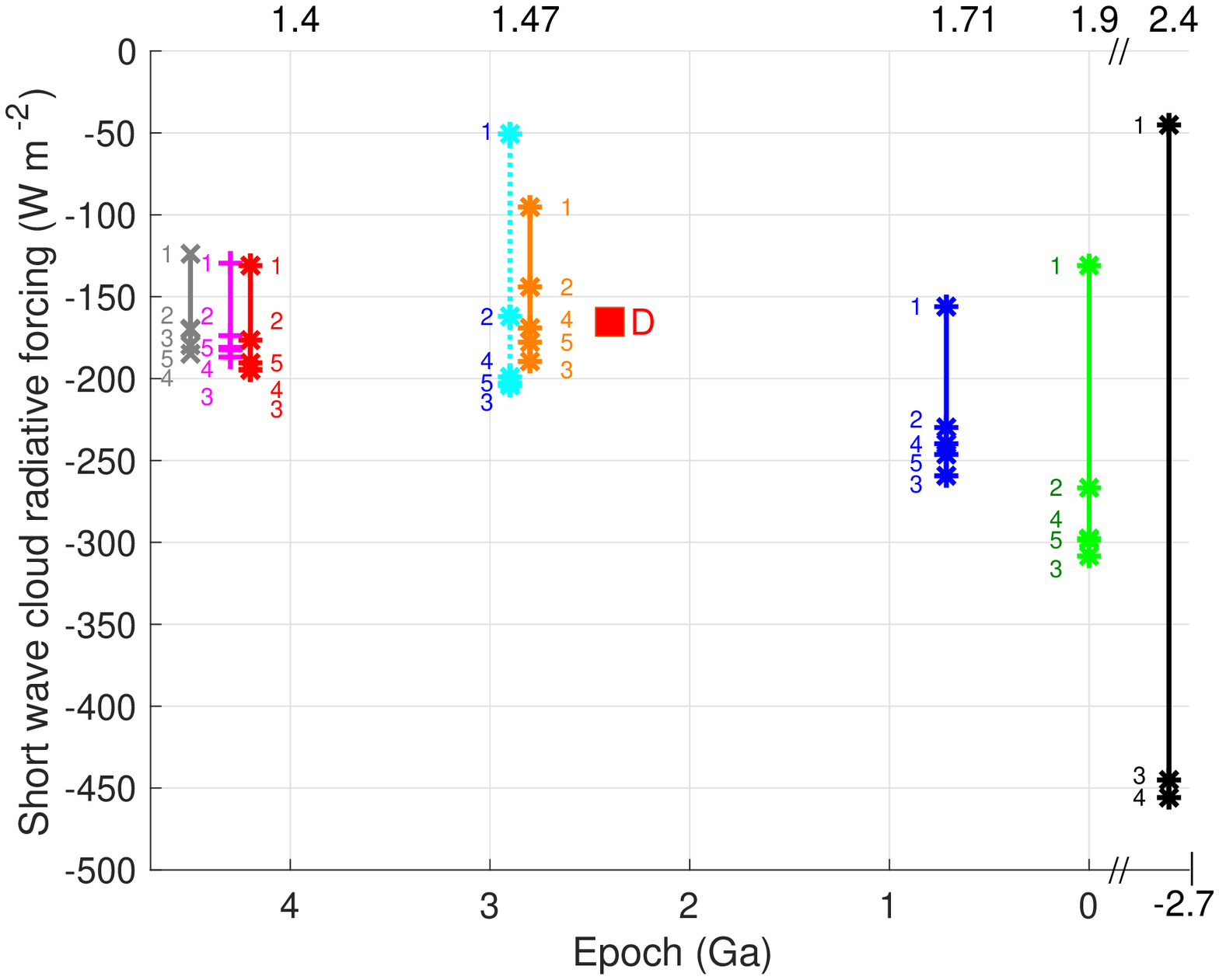} 
\caption{\small Shortwave cloud radiative forcing for the same evolutionary scenarios, defined as the difference between the solar radiation actually absorbed by the planet and how much would be absorbed if clouds were transparent. This is an estimate of how effective the clouds are at shielding the planet from the star's intense radiation.} \label{fig3}
\end{figure}

\begin{figure}[ht!] \centering
\includegraphics[width=\textwidth]{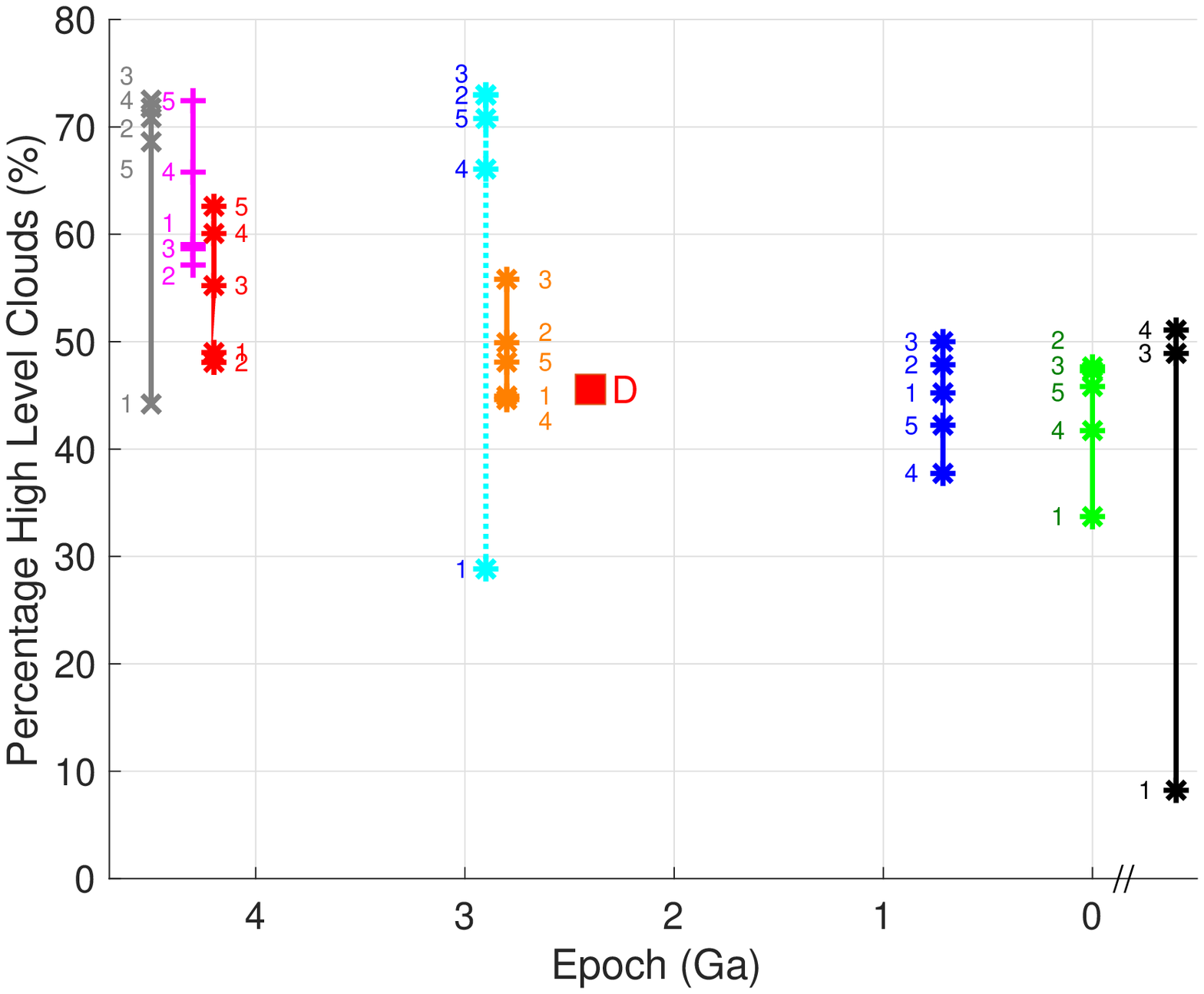} 
\caption{\small Percentage areal coverage of high level clouds (PCLDH). In \citeA{Way2018} it was shown that PCLDH plays a key role in the cloud albedo feedback for slow rotators.} \label{fig4}
\end{figure}

\subsection{4.2Ga}\label{subsection:4.2}
Since there are major uncertainties about what Venus' initial rotation rate was
(See Section \ref{sec:evolution}) we explored early post-magma ocean scenarios at 4.2Ga with three different retrograde initial rotation periods in the left hand part of Figures \ref{fig2},\ref{fig3},\ref{fig4}: --16 days (gray; experiments 6-10), --64 days (magenta; experiments 11-15) and --243 days (red; experiments 16-20). Each assumes a 1 bar CO$_2$-dominated atmosphere.  As one would expect from the studies of \citeA{Yang2014,Way2018} the faster spin rate simulations generally have higher temperatures because of the cloud processes discussed in those papers, but almost all of them reach equilibrium at a habitable global mean surface temperature. However, the clouds also differ to some degree because of water availability. Contrary to the work of \citeA{Abe2011,Kodama2019} the Arid-Venus cases all have higher surface temperatures than their counterparts. This is because those previous works used modern Earth's rotation rate, whereas the cloud processes on these slower rotating worlds better regulate the climate, more so the more water that is available for cloud formation. This analysis is backed up by Figure \ref{fig3} where we plot the shortwave cloud radiative forcing (SWCRF). The Arid-Venus simulations have the smallest (in magnitude) values, because a drier planet has less reflective clouds with less condensed water. In Figure \ref{fig4} we show the percentage of high level clouds (PCLDH), the dominant of the three cloud types (high, medium, low) in Table \ref{table:results}. Here the distinction between the Arid-Venus simulations and the others is not consistent across the different rotation periods, suggesting that in some cases middle and/or low level clouds make important contributions to SWCRF.

\subsection{2.9Ga}
\begin{figure}[ht!] \centering
\includegraphics[width=\textwidth]{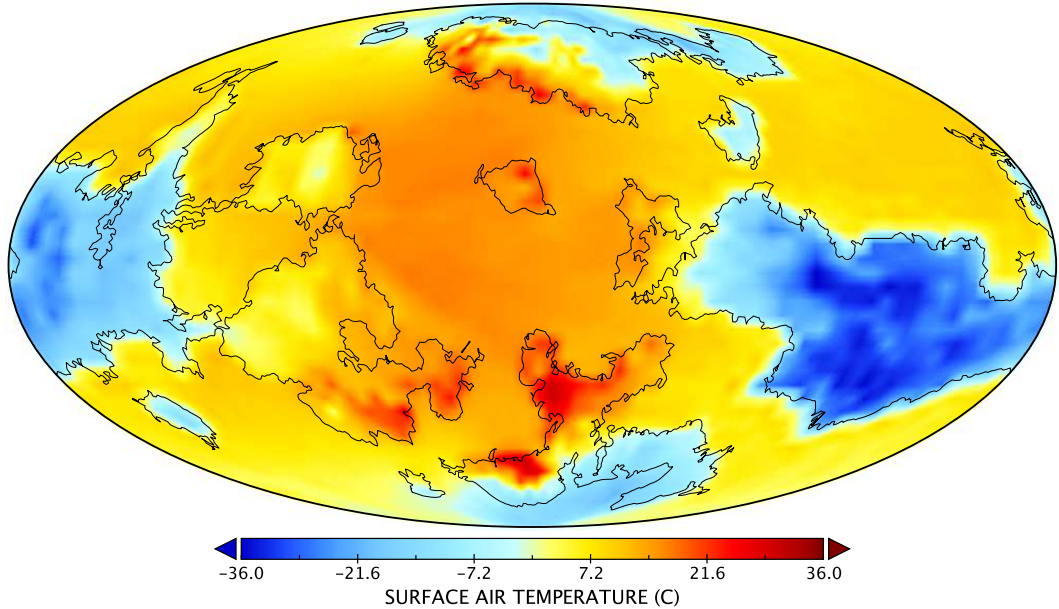} 
\caption{Simulation 28: mean surface temperature over 1/6th of a diurnal cycle. The substellar point is centered over the middle of the plot.} \label{fig5}
\end{figure}

Here we plot two different sets of simulations for N$_2$-dominated atmospheres: 1 bar (cyan; simulations 21-25) and 250mb (orange; simulations 26-30). In effect these portray representative possible atmospheres for an ancient Venus with liquid water that has evolved from an early CO$_2$-dominated atmosphere to a more Earth-like composition via the carbonate-silicate cycle feedback that is believed to regulate CO$_2$ on planets with liquid water. In both cases the Aqua-158m simulations have the highest mean surface temperatures with the Arid-Venus a close second, but all 10 simulations have moderate surface temperatures fairly similar to modern Earth. However as for simulations 6-20 the shortwave cloud radiative forcing is again the smallest for the Arid-Venus simulations (Figure \ref{fig3}) while also having less high cloud in (Figure \ref{fig4}) than the simulations with more surface water. Unsurprisingly, the thin 250mb atmospheres (simulations 26-30) have cooler surface temperatures in Figure \ref{fig2}. Simulations 27 \& 28 have lower mean surface temperatures than modern Earth. The surface temperature field for simulation 28 is plotted in Figure \ref{fig5} for reference. It exhibits fairly uniformly warm oceans, a signature of slowly rotating planets.  Continental temperatures are cooler, slightly below freezing on average, due to nighttime cooling that offsets daytime warming.
We also plot Simulation D from Paper I. The other 3 simulations from Paper I have similar values to their corresponding simulations herein. Simulation D is an N$_2$-dominated atmosphere, but is otherwise similar to ID 8 (a CO$_2$ dominated atmosphere) in Tables \ref{table:Experiments} and \ref{table:results}. It has a lower mean surface temperature than ID 8 as expected, but is significantly higher than the other simulations with larger rotation periods (ID=21--30) at 2.9Ga. It has less short wave cloud radiative forcing (Figure \ref{fig3}) and lower percentage of high level clouds (Figure \ref{fig4}) compared to the other 310m-Venus simulations at 2.9Ga. This is expected given its faster rotation period, stronger Coriolis force, and less contiguous clouds at the substellar point as discussed in Paper I.

\subsection{0.715Ga}
This epoch captures a possible final habitable phase on Venus, if the thick CO$_2$ atmosphere we see today was created by volcanic emissions during the global resurfacing event(s).  The spread in surface temperatures between simulations remains about the same as in previous epochs, but for all surface types the temperature is warmer than at 2.9 Ga because of the brighter Sun. Again, the Aqua-158m is has the highest surface temperatures with the Arid-Venus close behind as in the 2.9Ga epoch. The SWCRF is somewhat larger in general, due to both the stronger insolation and slightly reduced high cloud, but again the Arid-Venus has the smaller effect. The spread in high level clouds has shrunk considerably from the 1 bar simulations at 2.9Ga.

\subsection{Present Day}
This suite of simulations at Venus' present day insolation are designed to demonstrate that even under today's Sun the slow rotation cloud feedback effect would have remained strong as Venus' atmospheric pressure and composition remain unchanged. This points to the idea that it was not an increase in insolation that drastically changed Venus' clement climate of earlier epochs, but rather something else, which we speculate to be multiple/simultaneous large igneous provinces. The Arid-Venus simulations again have the highest temperatures and corresponding smallest SWCRF. This is more along lines of what we saw with the simulations at 4.2Ga, with climate forcing by a stronger Sun replacing climate forcing by a thicker greenhouse gas atmosphere as the primary reason for a warm climate.  

\subsection{Future}
Our last set of simulations at insolation values 2.4 times that of present day Earth are meant to show how long a temperate Venus-like world could have remained habitable for a given surface type. Our 10m-Venus and 310m-Earth simulations are not in equilibrium and so are not plotted. The Arid-Venus simulation is at nearly 100$^\circ$ Celsius. It appears to be approaching radiative equilibrium, but the simulation crashed after 20 years so it is difficult to be certain.  At this point the cloud/albedo feedback for the Arid-Venus case has decreased to Earth-like values (the SWCRF is a mere --50 W m$^{-2}$). This simulation has the lowest value of PCLDH, which makes it hard to counter the increased insolation at this time in order to keep mean surface temperatures below the boiling point of water. 

\subsection{General Trends}

A few relatively consistent trends are apparent from our simulations.
First, the Arid-Venus simulations tend to have the highest surface temperatures, smallest values of SWCRF and lowest percentages of PCLDH. In many cases the 10m-Venus simulations are next, but not always. This may point to the fact that even with 10m of available water the cloud/albedo feedback is generally effective at shielding this slowly rotating world from the intense solar radiation at all epochs when considering what modern Earth receives. The simulations with generally higher water availability and similar percentages of land-to-sea (310m-Venus and 310m-Earth) tend to cluster together in Figures \ref{fig2},\ref{fig3},\ref{fig4}. The 158m-Aqua simulations seem to float in between, perhaps because of the lower surface albedo in combination with the cloud albedo feedback.

\begin{figure}[ht!] \centering
\includegraphics[scale=0.9]{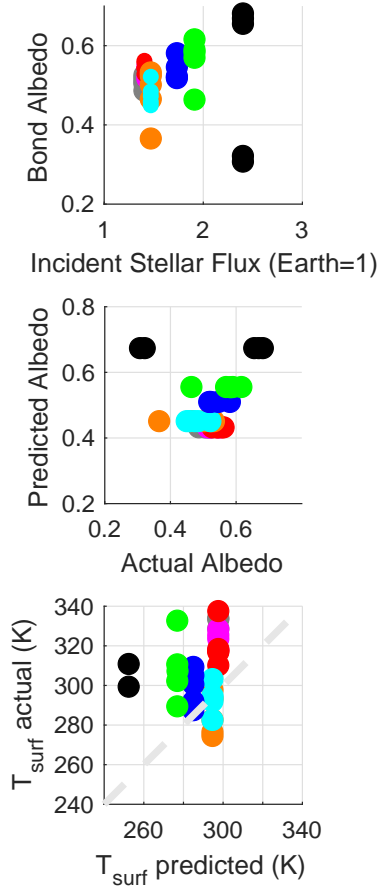} 
\caption{Upper panel: Bond albedo vs. incident solar flux for the planets in Figure 2. Middle panel: Predicted versus actual Bond albedos for the same set of planets after \citeA{DelGenio2019}.  Lower panel: Predicted vs. observed surface temperatures after \citeA{DelGenio2019}.} \label{fig6}
\end{figure}

Even without a transition to a runaway greenhouse, prior water loss due to the onset of a ``moist greenhouse" state may have been important to Venus' evolution.  A number of simulations in Table \ref{table:results} contain stratospheric water concentrations (Q$_{top}$) greater than 3$\times$10$^{-3}$ kg kg$^{-1}$, the traditional \citeA{Kasting1993b} limit for onset of the moist greenhouse. However, recent work by \citeA{Chen2019} for M-star planets has demonstrated that previous work may have overestimated water loss rates. Hence we should exercise more caution in using 3$\times$10$^{-3}$ kg kg$^{-1}$  as a hard value for the moist greenhouse until similar models are applied to G-star planets. Column Q$_{surf}$ is a check on the amount of water vapor at the surface of the model. \rocke{} runs with a fixed molecular mass at model start and ignores the spatially/temporally variable mass of water in calculating pressure gradients, so it is important to keep track of whether water becomes a non-negligible fraction of atmospheric mass (e.g., 20\% of the total or more) as the dynamics in the model will begin to be outside an acceptable range. Only in simulation 45 does this value go over the 20\% limit.

Figure \ref{fig6} shows calculations for what exoplanet astronomers might find for a population of ``exo-Venuses," some of them habitable and some not, in future observations. We use an ensemble of \rocke{} simulations of a variety of rocky planet types from which predictors for Bond albedo and surface temperature have been derived using insolation and star temperature as inputs \cite<>[hereafter DG19]{DelGenio2019}. Figure \ref{fig6} applies the predictor to our Venus evolutionary scenarios to determine the predictability of albedo and surface temperature. In general Bond albedo increases with insolation in the Venus simulations (upper panel), the exceptions being 4 of the Arid-Venus cases with limited surface water, fewer clouds and thus lower albedos than our other simulations. The DG19 predictor (middle panel) works well for all but these 4 cases, since it predicts a high cloud-controlled albedo for the wetter planets that have more and/or thicker clouds than others. For surface temperature (bottom panel), the predictor tends to underestimate the actual temperatures by roughly 20$^\circ$ or less in most cases, but by up to 50$^\circ$ for the hotter, drier, marginally habitable Arid-Venus cases.

\begin{figure}[ht!] \centering
\includegraphics[scale=0.4,angle=-90]{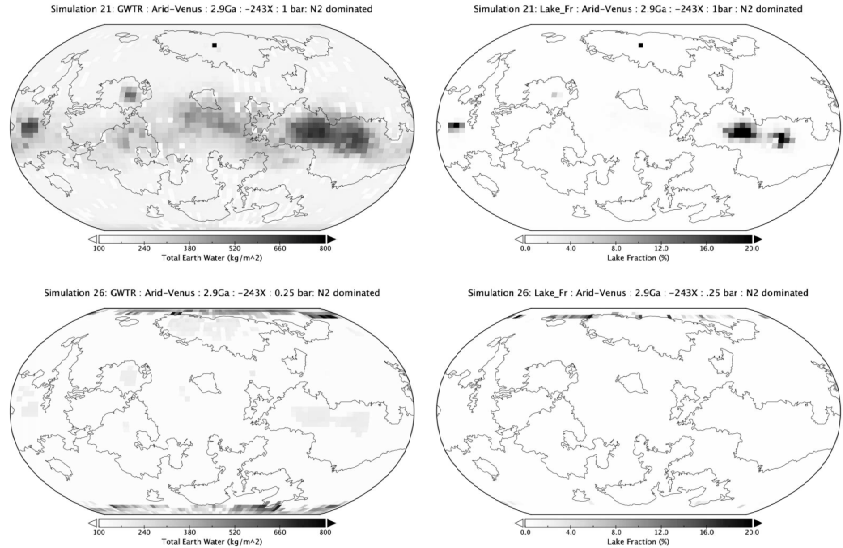}
\caption{\small GWTR left panels, LakeFR right panels for Top to Bottom: Simulation 21 and 26 (Arid-Venus).} \label{fig7}
\end{figure}

\begin{figure}[ht!] \centering
\includegraphics[scale=0.4,angle=-90]{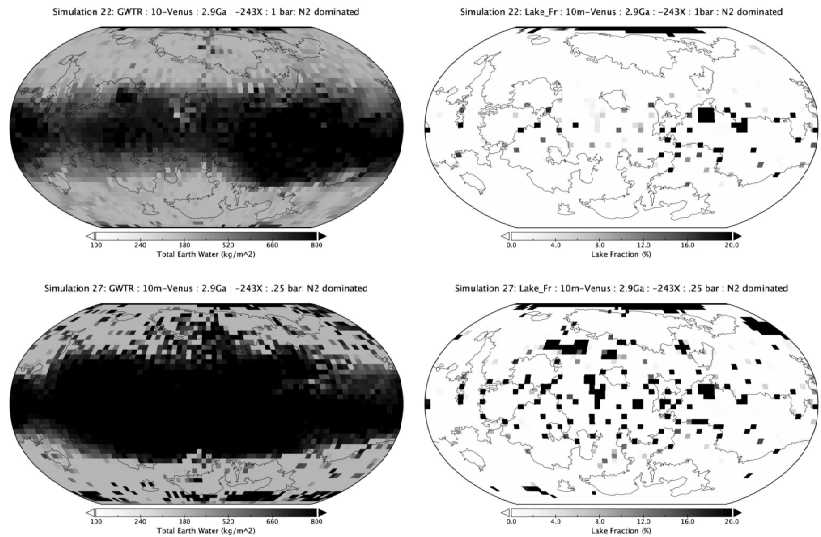}
\caption{\small GWTR left panels, LakeFR right panels for Top to Bottom: Simulation 22 and 27 (10m-Venus).} \label{fig8}
\end{figure}

Figure \ref{fig7} shows (left panels) the vertically integrated soil moisture and (right panels) lake fraction for two of our Arid-Venus simulations at 2.9 Ga with Earth-like atmospheres: Experiment 21 (1 bar, top panels) and Experiment 26 (250 mb, bottom panels).  The Arid-Venus cases are of particular interest because they are initialized with a spatially uniform subsurface soil water reservoir and no standing water bodies, and they then equilibrate to a heterogeneous distribution of surface and subsurface water depending on the climate and circulation and thus the local precipitation-evaporation competition.  The 1 bar atmosphere is typical of the behavior of most of the experiments in Table 1: In equilibrium, subsurface water collects primarily in the equatorial region where rising motion and precipitation is prevalent during the daytime, and especially in highland regions such as Aphrodite Terra. Lakes (which are not present in the initial condition) also form preferentially over the equatorial highlands.  The 250 mbar simulation is an outlier, with subsurface water and lakes arising primarily at high latitudes.  This is reminiscent of what is observed for the methane-ethane lakes on Titan, which \citeA{Mitchell2008} is able to reproduce in a GCM when a limited subsurface methane reservoir is assumed. 

Figure \ref{fig8} shows the same quantities for the analogous 10-m Venus simulations (Experiments 22 and 27).  These differ from the Arid-Venus cases not only because the planet contains more water, much of it in surface lakes rather than subsurface soil moisture, but also because the lakes fill the lowlands at the start of each simulation rather than being distributed uniformly across the planet.  With a larger water reservoir than that for the Arid-Venus planets, soil moisture collects throughout the tropics in the equilibrated climate, but still with a slight preference for the highland regions even though there is no standing water in the highlands in the initial condition.  But unlike the Arid-Venus planets, soil moisture also collects at the poles, both for the 1 bar and the 250 mb atmosphere planets.  Likewise, lakes in both simulations form in both the tropics and polar region, more so over the highlands for the 1 bar atmosphere but fairly uniformly distributed in longitude for the 250 mb atmosphere.

\section{Conclusion}\label{sec:conclusion}

Whether Venus' original water survived its initial MO stage, or whether significant water was delivered afterwards, is unknown.  It is therefore worth having a theoretical framework that considers the possibility of an early habitable Venus as a starting point for designing future observing strategies that might shed light on Venus' past. In this spirit, we envision the following possible climatic evolution for Venus and provide Figure \ref{fig9} as a guide:

\begin{figure}[ht!] \centering
\includegraphics[scale=0.7]{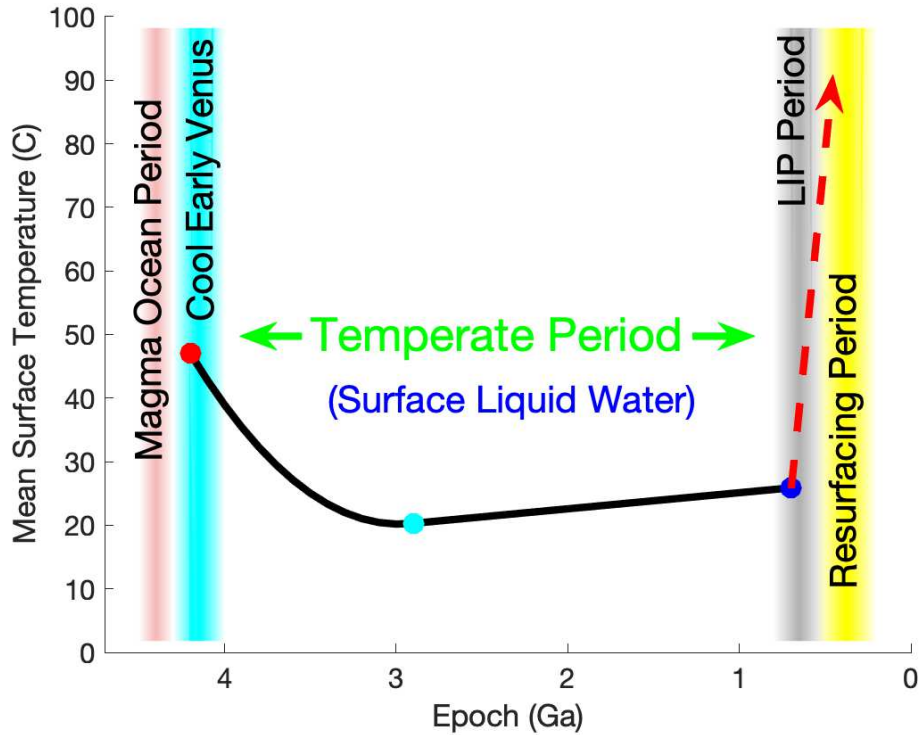} 
\caption{\small Graphical representation of Venus' possible climate history. The three data points represent the 1 bar atmospheres modeled at those points in time. The red dashed arrow to the right represents the transition to a moist/runaway greenhouse and eventually to Venus' present day surface temperature and atmospheric density.} \label{fig9}
\end{figure}

1.) Toward the end of the accretion period ($\sim$4.2Ga) Venus would have cooled rapidly as did Earth, as shown in the work of \citeA{Valley2002}. This would allow surface water to condense and early oceans to form.  Early oceans in turn could create significant tidal dissipation that would spin down Venus' rotation rate on a relatively short time scale as shown in work by \cite{Green2019} and described in Section \ref{sec:evolution} above. Solid body dissipation may have also been effective, see Section \ref{sec:evolution}. Slow rotation combined with an early ocean would provide the necessary ingredients for a dayside cloud deck to emerge, shielding the planet from high insolation and allowing at least some of the initial surface water to survive despite the planet being well inside the conventional inner edge of the habitable zone, as shown in the work of \citeA{Yang2014,Way2018}.

2.) The carbonate-silicate cycle in concert with interior volatile cycling would allow CO$_2$ draw-down while N$_2$ was outgassed, eventually reaching a balance producing an N$_2$-dominated atmosphere with trace amounts of CO$_2$ over gigayears with pressures ranging from several bars to hundreds of millibars. 

3.) We propose that any stable Venusian climate period came to an end at some period of time (e.g. 0.2Ga to 3 Ga) before the global resurfacing event. As mentioned previously surface age estimates for Venus range from as young as $\sim$180 Ma \cite{Bottke2016} to 750Ma \cite{McKinnon1997}.
We suggest that the ignition of multiple large igneous provinces (LIPs) became active around that time (over a period of 10s or 100s of millions of years). This would not have been the global LIP as proposed by \cite{Lopez1999} as it is not necessary and seems to have little support in the community. It is likely that the eruption of LIPs throughout Earth's history \cite{Ernst2014,Ernst2019} is a random stochastic process. This may imply that multiple {\emph large scale} LIPs have not occurred simultaneously on Earth by purely random chance, which is fortuitous for life as we know it today.  Venus may not have been as fortunate.  Unfortunately, little is known about Venus' interior structure today, much less its initial state and subsequent evolution, so the question remains as to whether deterministic evolutionary processes in the interiors of Venus-like planets will inevitably lead to catastrophic changes and uninhabitable end states. 

4.) Multiple large scale LIPs would have warmed the world markedly via the release of large amounts of CO$_2$ via one or more mechanisms and its greenhouse effect on the climate \cite<e.g.,>[]{Ogden2012} over 10s or 100s of millions of years.
It should be noted that degassing $\sim$ 90 bars worth of CO$_2$ via this mechanism would be difficult and is an open research topic.

5.) This warming could have hastened a movement toward the onset of a moist \cite{Kasting1993b} and possibly a runaway greenhouse state in which much of the surface water was lost via photodisassociation accompanied by hydrogen escape and an oxidized surface. Recall that present day Venus D/H measurements (if they are accurate) imply .6 to 16\% of Earth's present day surficial water stores \cite{Donahue1997}. Hence the timescale of total loss would be at least an order of magnitude faster than that described in \citeA{Kasting1993b}, although there is still uncertainty about the stratospheric conditions under which a moist greenhouse occurs \cite<e.g.,>[]{Chen2019} and when a runaway greenhouse is achieved \cite<e.g.,>[]{Kasting_Ackerman1986,Goldblatt2013,Ramirez2014}.
Work by \citeA{Grinspoon1993} on the D/H ratio may also support this hypothesis as they state: ``Thus (D/H)$_{obs}$ may be the isotopic signature of a catastrophic resurfacing in the past 0.5--1 Gyr."

6.) A major issue with any proposed evolutionary scenario with long lived surface water reservoirs then arises: what happened to the large quantities of oxygen expected to be left over from a Venusian ocean? This is the oxygen that would have remained in the atmosphere after the photo-dissociation of H$_{2}$O and the loss of the hydrogen \cite{Watson1981}, but not oxygen, via atmospheric escape. 
In fact large quantities of abiotically produced O$_{2}$ (100s to 1000s of bar) left over in such a scenario has been proposed as an observational signature in exoplanetary atmospheres on planets that have lost their oceans \cite{LugerBarnes2015}.
For our Arid-Venus scenario it may be possible to lose much of the oxygen via a combination of atmospheric escape \cite{Persson2018} and absorption by a surface like that of present day Venus \cite<e.g.,>[]{Gilmore2017}. However, recently submitted work
by \citeA{Persson2020} demonstrates that 0.3 m of a global equivalent layer of water could have been lost via atmospheric escape alone in the past $\sim$4Ga. Hence the 0.2 m global equivalent layer of water in our Arid-Venus scenario fits within this framework.
Yet work by \citeA{Persson2018,Masunaga2019,Persson2020} and their estimates for O+ escape rates shows that atmospheric escape alone is not sufficient to remove larger reservoirs of oxygen left over from the oceans in our other non-Arid-Venus simulations. 
It should be noted that the escape estimates from \citeA{Persson2020} are distinctly lower than previous work by \citeA{Chassefiere1996,Chassefiere1997} where they are mainly concerned with escape during a more active younger sun. The work of \citeA{Abe2011} also gives much higher escape rates estimating that an entire Earth Ocean's volume could be lost in 600Myr to 14Myr depending on how active the sun is.
Some caveats go with this work in that the Venus atmosphere simulated for our temperate period (more akin to an Archean Earth atmosphere than modern day Earth or Venus) is N$_2$ dominated with 400ppmv CO$_2$ and 1ppmv CH$_4$ and is very different from that of modern Venus with thermospheric and exospheric temperatures likely to be distinct and possibly affecting escape rates \cite{Airapetian2017}. 

We propose that the large-scale resurfacing evident on Venus today, which took place over 100s of millions of years, is a possible effective answer. It must be noted that this would be separate from the earlier (in geological time) LIP scenario above.
Since the Magellan mission it has been known that ~80\% of the surface of Venus is relatively young, with estimates ranging from 300-700$Myr$ old as mentioned in Section 4. These newly exposed basalts would be the ideal sink for large quantities of oxygen (possibly 100s of bars) over 100s of millions of years. According to \cite{Lecuyer2000} Venus would need to oxydize a rock layer $\sim$50 km deep to absorb an Earth Ocean's worth of oxygen, and they propose a mechanism for doing so while citing the earlier work of \citeA{Pieters1986}. Note that none of the oceans proposed herein are close to an Earth's ocean in volume, hence the number could be much smaller. For example, work by \citeA{Grinspoon1993} and \citeA{Head1992} note that the volume of magma necessary to cover all pre-existing craters would need to be a global layer $\sim$10 km deep, and that would be sufficient for the volume of oceans proposed in our models.
Some fraction of the oxygen may actually be deep within Venus’ lithosphere and possibly even within its mantle. This may be consistent with coronae-related subduction hypotheses \cite{Sandwell1992,Davaille2017} and other ideas about downwelling-associated highlands \cite<e.g.>[see Fig 2]{Head1992}, where thicker crustal regions may exhibit orogenesis \cite<e.g.,>[]{Head1990} and sinking of parts of the lithosphere into the mantle \cite<e.g.,>[]{Lenardic1991,Bindschadler1992}.


7.) The loss of water would in turn change the planet from an initial subductive or mobile plate tectonic mode to more of a stagnant lid mode (as on present day Venus \& Mars) since it is currently believed that water plays a key role in plate tectonics on Earth \cite<e.g.,>[]{Grove2012,Lecuyer2014}. This scenario fits in very nicely with the recent work of \cite{WellerKiefer2019} who give a timescale of order 1Gyr for the transition from a mobile to a stagnant lid mode on Venus in their simplified model. Without a mechanism to efficiently cycle volatiles in a stagnant lid mode \cite<e.g.,>[]{Tosi2017,Honing2019}, outgassing would have continued without the major weathering and subduction surface sinks that operate on Earth, hence CO$_2$ and N$_2$ would build up over time to reach the levels we see on Venus today. Some studies have also shown that even in a stagnant lid mode it is possible to cycle volatiles, possibly up to gigayears in time \cite<e.g.,>[]{Foley_Smye2018,Godolt2019}, but these mechanisms depend on the initial CO$_2$ budget and the retention of at least some water after cooldown. 

8.) This stagnant lid mode may then allow very large mantle upwelling and/or downwelling centers that would produce some of the features we see on Venus' surface today produced over hundreds of millions of years, as described most recently in the works of \citeA<e.g>[]{Rolf2018,WellerKiefer2019}.

Our scenario can also fit within the Fortunian, Guineverian \& Atlian periods proposed in the works of \citeA{IvanovHead2015b} and \citeA{Airey2017}, as it is not possible to constrain the start of the LIP period we propose with the data we have today. Our LIP period could easily have concluded in the pre-Fortunian or Fortunian period $\sim$1.5Ga.

One of the remaining quandaries in our hypothesis is the fact that the 92 bar atmosphere we see on Venus today must go back at least as far as the age of the visible surface because there are fewer small craters ($<$35 km in diameter) to be seen in the Magellan data \cite<>[see Figure 2]{Schaber1992}. Certainly smaller craters would be visible if the atmosphere had been significantly thinner in the lifetime of the observed surface when atmospheric filtering of smaller impactors would have been less prevalent.

For example, assuming the tesserae are the oldest stratigraphic units why are there no small craters present if the present day atmosphere is not a primordial feature from many eons ago? One resolution to this problem could be that the tesserae are not as old as we think they are, and until we date these units and the basaltic flats we really won't know. Secondly, as mentioned above, there is the possibility that there have been multiple resurfacing events and the tesserae are left-over from one of the previous events. Neither is a terribly optimistic scenario if one is hoping that some of tesserae may be remnant crust from a period of hosting surface water. Finally, a large impactor may be the cause of the ‘catastrophic’ climate change we propose. This could have also played a role in resetting the clock on the surface of Venus reconciling the lack of small craters. In this scenario the LIP hypothesis plays a partial role in the evolution of Venus' climate. This would be similar to what we have seen in the on-going debate regarding the K-Pg period on Earth  \cite<e.g.>[]{Hull2020,Schoene2019,Sprain2019}.
Such an impactor's imprint would have long been lost due to Venus' relatively young surface. Our comments about impactors in Section \ref{section:surface-history} apply here as well.

Clearly a great deal more modeling work and more observations are required to confirm or refute this hypothesis. Did Venus follow the 'canonical' path with Earth-like conditions in it's early history and then experience a moist-runaway greenhouse due to increasing solar insolation?
Did it experience a longer period of habitability throughout most of its history, with its demise and present hothouse state the consequence of a series of LIP events releasing CO$_2$ bound up in the crust as on Earth, and/or released from the deep interior where CO$_2$ is more easily sequestered \cite<e.g.,>{Kuramoto1996}? Or did it become bone dry in an extended magma ocean phase in the first 100$Myr$, as described in \citeA{Hamano2013} for Type II planets? 

We believe the only scenario we can begin to rule out with the present work is the `canonical path' since there is no evidence that an early period of habitability would have been affected by increasing solar luminosity in the first billion years. In essence, if Venus had habitable surface conditions with surface liquid water $\sim$4Ga then the same cloud albedo effect that allows such a scenario would continue for eons. On the other hand we will not be able to distinguish between the two remaining scenarios until we return to Venus to make proper noble gas and other elemental and isotopic measurements at the surface \cite{Baines2013} and better constrain escape processes at the top of the atmosphere through time. The latter will also rely upon how such gases escape from present day Earth given the possibility that Venus may have had a magnetic field in previous epochs, even if it is not clear how important the magnetic field is to escape processes in general \cite{Gunell2018}.  Likewise, whether the actual evolution of the one Venus we can visit is the ultimate fate of all highly irradiated rocky planets or an accident of an evolutionary path that might have proceeded differently in other circumstances \cite<e.g.>[]{Lenardic2016} is not known.  The stakes are high for answering this question, since many exoplanets have been discovered in the ``Venus zone" just inside the traditional inner edge of the habitable zones of other stars \cite{Kane2014}.  Efforts to simultaneously characterize the CO$_2$ concentrations and climates of a number of these exoplanets, combined with a focused observational strategy for unveiling the history of the ``exoplanet next door" to Earth in our own solar system \cite{Kane2019}, will be our best chance to understand whether the envelope for habitability and the emergence of life is much broader than usually assumed. 

\begin{table}[ht!]
\caption{Results}
\label{table:results}
\scriptsize
\begin{tabular}{|l|r|r|r|c|r|c|c|c|c|c|} 
\hline
ID$^{a}$&  Runtime & Temp$^{b}$& Balance$^{c}$  & Q$_{top}^{d}$&Q$_{surf}^{e}$& Albedo   & Albedo  & Clouds & Clouds & Clouds\\
  &  \multicolumn{1}{c|}{years}   & \multicolumn{1}{c|}{C}   &W m$^{-2}$& kg/kg    &  \multicolumn{1}{c|}{\%}      &Planetary & Surface & High & Medium & Low\\
\hline
01 &      71 &  262 &    0.01 & 1.36e-05 &   0.08 &      54 &   29 &      54 &       5 &       0 \\ 
02 &     111 &  151 &   27.21 & 2.67e-05 &   3.74 &      54 &   18 &      33 &      17 &      33 \\ 
03 &      67 &  121 &   21.31 & 4.50e-04 &   7.87 &      52 &   11 &      45 &      23 &      37 \\ 
04 &      52 &  120 &   51.28 & 1.74e-04 &  10.01 &      44 &    7 &      36 &      17 &      51 \\ 
05 &      56 &  123 &   36.27 & 4.28e-04 &   9.16 &      49 &   10 &      46 &      24 &      44 \\ 
\hline
06/{\color{gray}\bf01} &     100 &   96 &    0.06 & 1.73e-03 &   1.65 &      51 &   29 &      44 &       2 &       0 \\ 
07/{\color{gray}\bf02} &     100 &   76 &    0.41 & 9.34e-03 &   7.70 &      53 &   16 &      71 &      32 &      21 \\ 
08/{\color{gray}\bf03} &      53 &   61 &    5.92 & 6.49e-03 &   7.50 &      50 &   10 &      72 &      32 &      47 \\ 
09/{\color{gray}\bf04} &     100 &   59 &   -0.48 & 6.23e-03 &   6.95 &      52 &    7 &      73 &      35 &      62 \\ 
10/{\color{gray}\bf05} &      60 &   60 &    1.90 & 5.46e-03 &   7.19 &      52 &    9 &      69 &      37 &      53 \\ 
\hline
11/{\color{magenta}\bf01} &     200 &   73 &    0.05 & 9.03e-04 &   1.70 &      52 &   29 &      59 &       4 &       1 \\ 
12/{\color{magenta}\bf02} &     500 &   50 &    0.04 & 5.94e-04 &   2.07 &      55 &   16 &      57 &      10 &      10 \\ 
13/{\color{magenta}\bf03} &     200 &   45 &    0.87 & 1.56e-03 &   3.15 &      54 &   10 &      59 &      18 &      43 \\ 
14/{\color{magenta}\bf04} &     156 &   53 &    0.11 & 3.47e-03 &   4.81 &      51 &    7 &      66 &      25 &      60 \\ 
15/{\color{magenta}\bf05} &     200 &   56 &    0.44 & 4.58e-03 &   5.65 &      52 &    9 &      72 &      28 &      49 \\ 
\hline
16/{\color{red}\bf01} &     300 &   64 &    0.01 & 8.34e-04 &   0.80 &      53 &   28 &      49 &       4 &       1 \\ 
17/{\color{red}\bf02} &     500 &   44 &    0.06 & 5.68e-04 &   1.40 &      56 &   16 &      48 &       9 &       8 \\ 
18/{\color{red}\bf03} &    1000 &   37 &    0.13 & 8.18e-04 &   2.07 &      56 &   10 &      55 &      15 &      38 \\ 
19/{\color{red}\bf04} &     500 &   44 &   -0.50 & 1.59e-03 &   3.01 &      54 &    7 &      60 &      17 &      56 \\ 
20/{\color{red}\bf05} &     500 &   45 &    0.23 & 1.88e-03 &   3.15 &      54 &    9 &      63 &      19 &      44 \\ 
\hline
21/{\color{cyan}\bf01} &     500 &   22 &    0.02 & 2.52e-05 &   0.36 &      46 &   28 &      45 &       6 &       4 \\ 
22/{\color{cyan}\bf02} &    3000 &   21 &   -0.04 & 3.08e-05 &   0.61 &      47 &   17 &      50 &      10 &      12 \\ 
23/{\color{cyan}\bf03} &    1000 &   10 &   -0.05 & 2.53e-06 &   0.70 &      52 &   10 &      56 &      20 &      48 \\ 
24/{\color{cyan}\bf04} &     500 &   30 &   -0.12 & 3.51e-05 &   2.32 &      45 &    6 &      45 &      14 &      37 \\ 
25/{\color{cyan}\bf05} &    1000 &   19 &    0.05 & 5.87e-06 &   1.24 &      48 &    9 &      48 &      22 &      39 \\ 
\hline
26/{\color{orange}\bf01} &     300 &   20 &    0.11 & 5.94e-03 &   0.53 &      37 &   29 &      29 &       5 &       9 \\ 
27/{\color{orange}\bf02} &    1000 &    8 &   -0.14 & 9.98e-03 &   1.26 &      50 &   20 &      73 &      17 &      27 \\ 
28/{\color{orange}\bf03} &    1000 &    3 &   -0.02 & 6.15e-03 &   1.94 &      52 &   13 &      73 &       8 &      13 \\ 
29/{\color{orange}\bf04} &    1000 &   25 &    0.34 & 2.69e-02 &   6.39 &      47 &    6 &      66 &       1 &       0 \\ 
30/{\color{orange}\bf05} &     500 &   10 &    0.08 & 8.58e-03 &   2.76 &      50 &   10 &      71 &       7 &      10 \\ 
\hline
31/{\color{blue}\bf01} &     300 &   32 &   -0.03 & 1.12e-04 &   0.63 &      52 &   28 &      45 &       7 &       4 \\ 
32/{\color{blue}\bf02} &     297 &   19 &    0.02 & 8.32e-05 &   0.87 &       2 &    1 &      48 &      20 &      26 \\ 
33/{\color{blue}\bf03} &    1000 &   15 &   -0.03 & 1.18e-05 &   0.95 &      58 &   10 &      50 &      28 &      47 \\ 
34/{\color{blue}\bf04} &     500 &   36 &   -0.03 & 6.41e-04 &   3.32 &      52 &    7 &      38 &      17 &      42 \\ 
35/{\color{blue}\bf05} &     500 &   28 &    0.27 & 1.11e-04 &   2.04 &      55 &    9 &      42 &      26 &      40 \\ 
\hline
36/{\color{green}\bf01} &      50 &   60 &   -0.09 & 3.05e-03 &   0.58 &      46 &   29 &      34 &       2 &       1 \\ 
37/{\color{green}\bf02} &    2000 &   29 &   -0.01 & 4.33e-04 &   1.17 &      59 &   17 &      48 &      16 &      23 \\ 
38/{\color{green}\bf03} &    1000 &   16 &    0.12 & 3.52e-05 &   1.08 &      62 &   10 &      47 &      33 &      47 \\ 
39/{\color{green}\bf04} &     500 &   38 &   -0.07 & 1.37e-03 &   3.57 &      57 &    7 &      42 &      21 &      42 \\ 
40/{\color{green}\bf05} &    1000 &   34 &    0.35 & 1.10e-03 &   2.92 &      58 &    9 &      46 &      24 &      41 \\ 
\hline
41/{\bf01} &      20 &   99 &    0.52 & 3.13e-03 &   0.43 &      32 &   29 &       8 &       0 &       0 \\ 
42/{\bf02} &       8 &  288 &   90.21 & 1.00e-01 &  11.71 &      31 &   17 &      31 &       0 &       0 \\ 
43/{\bf03} &    1000 &   26 &    0.01 & 3.35e-03 &   1.99 &      68 &   11 &      49 &      33 &      54 \\ 
44/{\bf04} &     500 &   38 &    0.23 & 5.52e-03 &   3.75 &      67 &    8 &      51 &      44 &      54 \\ 
45/{\bf05} &      35 &   83 &   32.64 & 1.05e-01 &  30.36 &      66 &   10 &      89 &      19 &      16 \\ 
\hline
D          &    2000 &   56 &    0.19 & 5.31e-03 &   8.97 &      44 &   10 &      46 &      39 &      45 \\ 
\hline
\end{tabular}
\vspace{0.1cm}
\newline
$^a$ID: The colored numbers correspond to those in Figure \ref{fig2}.\newline
$^b$Temp: Surface temperature in Celsius.\newline
$^c$Balance: These numbers come from 50 year averages unless the Runtime (Column 2) is less than 150 years, then the average is 10 years.\newline
$^d$Specific humidity in top layer of the atmosphere.\newline
$^e$Surface humidity as percentage of atmosphere.\newline

\end{table}


%
%
%
%
\newpage
\appendix
\section{Energy balance and temperature}\label{appendix:SI}

Different simulations reach radiative balance sooner than others, 
while some never reach it at all.
Herein we plot the energy balance (in units of W m$^{-2}$) and surface temperature (in Celsius) as a function of simulation year. This should allow the reader to have a better grasp of which simulations are appropriate for a given interest.

\begin{figure}[ht!] \centering
\includegraphics[scale=0.55,angle=-90]{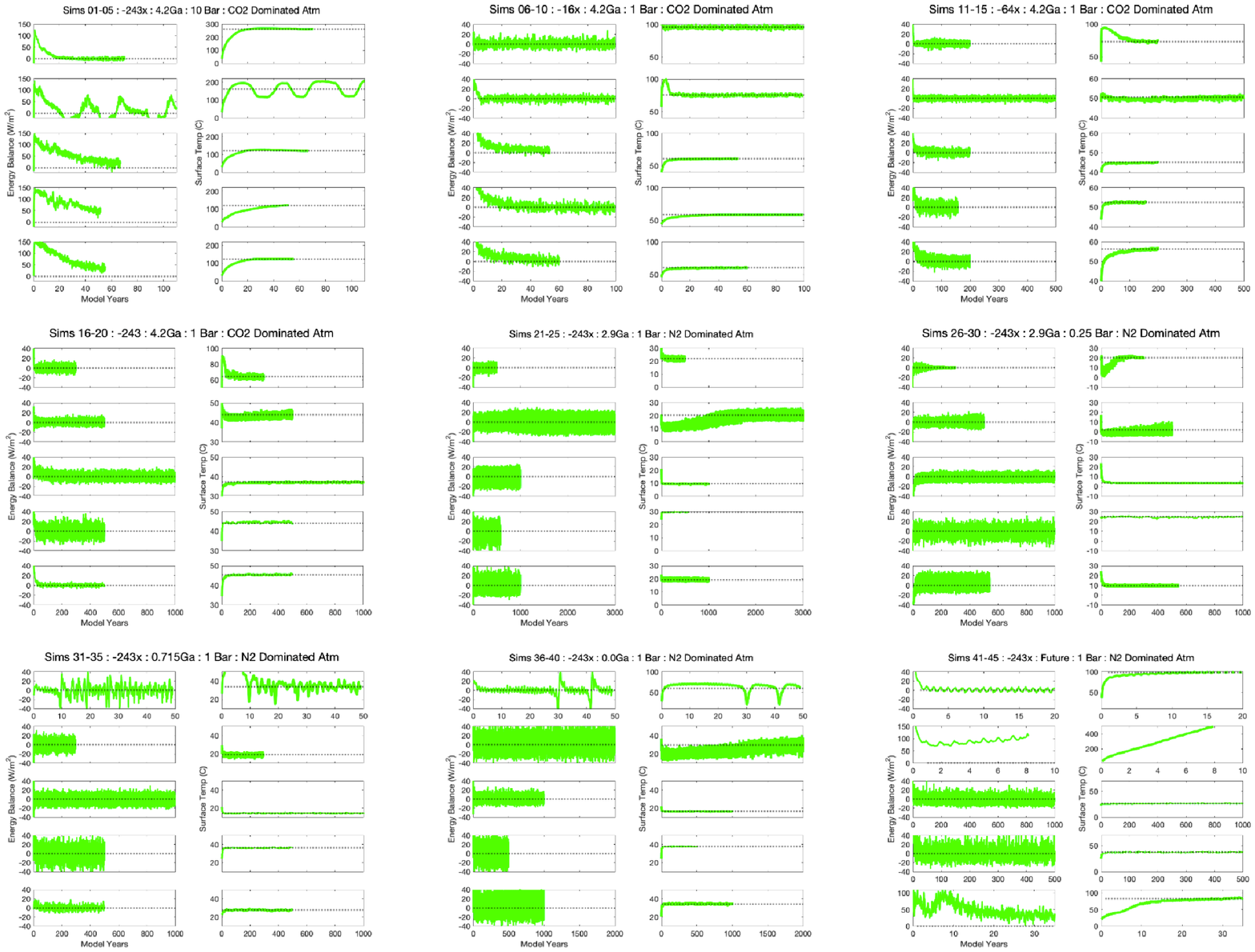}
\caption{\small Energy Balance (also called Net Radiative Balance) in the left columns and surface temperature in the right columns as a function of simulation year for all simulations in this study. Note that not all limits on the x or y axes are the same.} \label{fig10}
\end{figure}

\acknowledgments
This work was supported by the NASA Astrobiology Program through collaborations arising from our participation in the Nexus for Exoplanet System Science (NExSS). Resources supporting this work were provided by the NASA High-End Computing (HEC) Program through the NASA Center for Climate Simulation (NCCS) at Goddard Space Flight Center.
MJW acknowledges support from the GSFC Sellers Exoplanet Environments Collaboration (SEEC), which is funded by the NASA Planetary Science Division’s Internal Scientist Funding Model.
Thanks goes to Rory Barnes for his help with his eqTide program and Eric Wolf for his discussions involving HITRAN and HITEMP.
Thanks also to Wade Henning,Tony Dobrovolskis, and Kevin Zahnle for their thoughts on spin evolution. This paper also benefited from useful discussions with Mareike Godolt, Lena Noack, Richard Ernst, Linda Sohl, Matthew Weller, Christine Houser, Kostas Tsigaridis, Igor Aleinov, and Glyn Collinson. This paper also benefited greatly from the detailed comments of three referees: Jim Kasting, David Grinspoon and Cedric Gillmann.
All NetCDF data used in this publication can be downloaded from the Zenodo open access data portal:\newline
http://doi.org/10.5281/zenodo.3707251


%
%

\bibliography{bibliographyAGU}

%
%
%
%

\end{document}